\documentclass[sigconf]{acmart}

\usepackage{multirow}
\usepackage{color}
\usepackage{colortbl}
\usepackage{stfloats}
\usepackage{subcaption}
\usepackage{enumitem}
\usepackage{graphicx}
\usepackage{pifont}
\usepackage{tabularx}
\usepackage{longfbox}
\usepackage{soul}
\usepackage{tikz}
\usepackage{etoolbox}

\makeatletter
\newdimen\@tempdimd
\makeatother

\newfboxstyle{boxparam}{padding=1.5pt, padding-left=2pt, padding-right=2pt, height=7.5pt, background-color=lightgray, border-radius=2pt, border-right-width=1pt, border-bottom-width=1pt, border-color=gray}

\definecolor{selectionboxcolor}{HTML}{f0f0f0}

\newfboxstyle{selectionboxparam}{boxparam, background-color=selectionboxcolor}

\newcommand{\selectionbox}[1]{\lfbox[selectionboxparam]{\textsf{#1}}}

\newcommand{\rectwrapsmall}[2]{\lfbox[boxparam, border-radius=0pt, padding-left=2pt, padding-right=2pt, height=5.5pt, border-width=0pt, background-color=#1]{\sffamily{\textcolor{white}{#2}}}}

\newcommand{\blackrectsmall}[1]{\rectwrapsmall{darkgray}{#1}}

\definecolor{fillgreen}{HTML}{B6E37C}

\definecolor{quotebackground}{HTML}{EFEFEF}
\definecolor{tableheader}{HTML}{EFEFEF}
\definecolor{tablegrayline}{HTML}{e0e0e0}

\newcommand{\sysname}{\textsc{Autiverse}}

\renewcommand{\checkmark}{\ding{51}}

\definecolor{preparationcolor}{HTML}{d1476f}
\definecolor{articulationcolor}{HTML}{f37778}
\definecolor{verificationcolor}{HTML}{ffb000}
\definecolor{elaborationcolor}{HTML}{75ad57}
\definecolor{revisioncolor}{HTML}{66a8e2}
\definecolor{wrapupcolor}{HTML}{40539b}

\newfboxstyle{phaseboxparam}{padding=2pt, padding-left=4pt, padding-right=4pt, border-style=none, height=6pt, border-radius=5pt}

\newcommand{\phasepreparation}{\lfbox[phaseboxparam, background-color=preparationcolor]{\sffamily{\textcolor{white}{Preparation}}}}

\newcommand{\phasearticulation}{\lfbox[phaseboxparam, background-color=articulationcolor]{\sffamily{\textcolor{white}{Articulation}}}}

\newcommand{\phaseverification}{\lfbox[phaseboxparam, background-color=verificationcolor]{\sffamily{\textcolor{white}{Verification}}}}

\newcommand{\phaseelaboration}{\lfbox[phaseboxparam, background-color=elaborationcolor]{\sffamily{\textcolor{white}{Elaboration}}}}

\newcommand{\phaserevision}{\lfbox[phaseboxparam, background-color=revisioncolor]{\sffamily{\textcolor{white}{Revision}}}}

\newcommand{\phasewrapup}{\lfbox[phaseboxparam, background-color=wrapupcolor]{\sffamily{\textcolor{white}{Wrapup}}}}

\newcommand{\eg}{\textit{e.g.}}
\newcommand{\ie}{\textit{i.e.}}
\newcommand{\cf}{\textit{c.f.}}

\newcommand{\labelphantom}[1]{%
  \parbox{0pt}{\phantomsubcaption\label{#1}}%
}

\definecolor{parentcolor}{HTML}{cfe5cc}
\definecolor{childcolor}{HTML}{edd8b9}

\newcommand{\parentwrap}[1]{\lfbox[boxparam, border-width=0pt, background-color=parentcolor]{#1}}

\newcommand{\childwrap}[1]{\lfbox[boxparam, border-width=0pt, background-color=childcolor]{#1}}

\newcommand{\parent}[1]{\parentwrap{P#1}}

\newcommand{\child}[1]{\childwrap{C#1}}

\DeclareRobustCommand{\needtocheck}[1]{\textcolor{red}{#1}}

\newtoggle{clean}
\toggletrue{clean} 

\DeclareRobustCommand{\revised}[1]{%
  \iftoggle{clean}{#1}{\textcolor{blue}{#1}}%
}

\newtoggle{cameraclean}
\toggletrue{cameraclean} 

\DeclareRobustCommand{\cameraready}[1]{%
  \iftoggle{cameraclean}{#1}{\textcolor{red}{#1}}%
}

\newenvironment{revisedenv}{%
    \begingroup
    \iftoggle{clean}{}{%
        \color{blue}%
    }%
    \ignorespaces
}{%
    \endgroup
    \ignorespacesafterend
}

\newcommand{\deleted}[1]{%
  \ifthenelse{\boolean{clean}}{}{%
    \textcolor{cyan}{\st{#1}}%
  }%
}

\newcommand{\deletedsubsection}[1]{%
  \ifthenelse{\boolean{clean}}{}{%
    \subsection{\textcolor{cyan}{[Deleted] #1}}%
  }%
}

\newcommand{\circledigit}[1]{\textbf{\normalsize{\textsf{\textcircled{\footnotesize{#1}}}}}}

\newcommand{\ipstart}[1]{\vspace{1mm} \noindent{\textbf{\textit{#1.}}}}

\newcommand{\sourcecode}{\urlstyle{tt}\url{https://naver-ai.github.io/autiverse}}

\AtBeginDocument{%
  }

\copyrightyear{2026}
\acmYear{2026}
\setcopyright{cc}
\setcctype{by-nc-nd}
\acmConference[CHI '26]{Proceedings of the 2026 CHI Conference on Human Factors in Computing Systems}{April 13--17, 2026}{Barcelona, Spain}
\acmBooktitle{Proceedings of the 2026 CHI Conference on Human Factors in Computing Systems (CHI '26), April 13--17, 2026, Barcelona, Spain}
\acmPrice{}
\acmDOI{10.1145/3772318.3791381}
\acmISBN{979-8-4007-2278-3/2026/04}

\begin{document}

\title{\sysname{}: Eliciting Autistic Adolescents' Daily Narratives through AI-guided Multimodal Journaling}

\settopmatter{authorsperrow=0}

\author{Migyeong Yang}
\authornote{Migyeong Yang conducted this work as a research intern at NAVER AI Lab while at Sungkyunkwan University.}
\orcid{0000-0002-3569-9558}
\affiliation{%
  \institution{NAVER AI Lab}
  \country{Republic of Korea}}
\email{migyeong.yang@navercorp.com}

\author{Kyungah Lee}
\orcid{0009-0009-7872-6454}
\affiliation{%
  \institution{Dodakim Child Development Center}
  \country{Republic of Korea}
}
\email{hiroo6900@hanmail.net}

\author{Jinyoung Han}
\orcid{0000-0002-8911-2791}
\affiliation{%
  \institution{Sungkyunkwan University}
  \country{Republic of Korea}}
\email{jinyounghan@skku.edu}

\author{SoHyun Park}
\orcid{0000-0001-8703-0584}
\affiliation{%
  \institution{NAVER Cloud}
  \country{Republic of Korea}}
\email{sohyun@snu.ac.kr}

\author{Young-Ho Kim}
\orcid{0000-0002-2681-2774}
\affiliation{%
  \institution{NAVER AI Lab}
  \country{Republic of Korea}
}
\email{yghokim@younghokim.net}

\begin{abstract}
Journaling can potentially serve as an effective method for autistic adolescents to improve narrative skills. However, its text-centric nature and high executive functioning demands present barriers to practice.
We present \sysname{}, an AI-guided multimodal journaling app for tablets that scaffolds daily narratives through conversational prompts and visual supports. \sysname{} elicits key details of an adolescent-selected event through a stepwise dialogue with peer-like, customizable AI and composes them into an editable four-panel comic strip.
Through a two-week deployment study with 10 autistic adolescent-parent dyads, we examine how \sysname{} supports autistic adolescents to organize their daily experience and emotion.
\revised{Our findings show \sysname{} scaffolded adolescents' coherent narratives, while enabling parents to learn additional details of their child's events and emotions. Moreover, the customized AI peer created a comfortable space for sharing, fostering enjoyment and a strong sense of agency. Drawing on these results, we discuss implications for adaptive scaffolding across autism profiles, socio-emotionally appropriate AI peer design, and balancing autonomy with parental involvement.}

\end{abstract}



\begin{teaserfigure}
  \includegraphics[width=\textwidth]{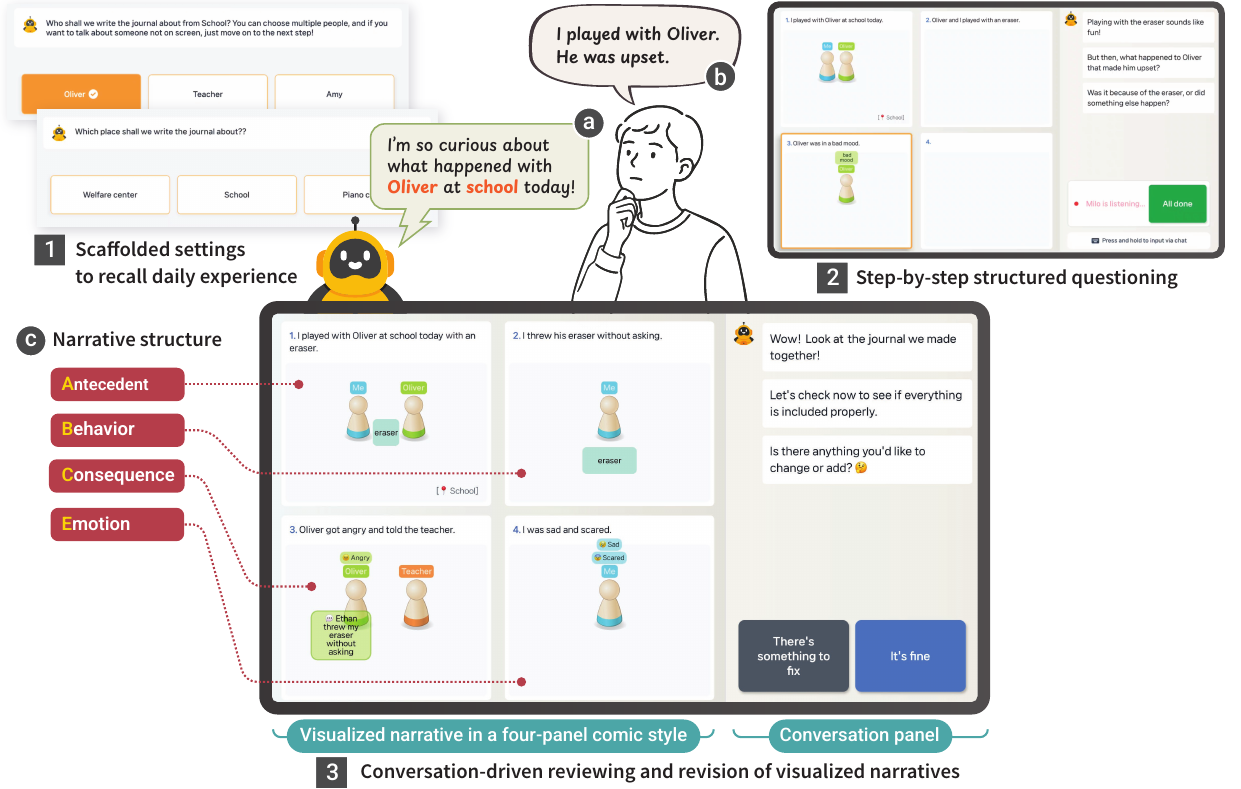}
  \caption{\sysname{} enables autistic adolescents' journaling by providing scaffolding layers driven by a peer AI. A journaling session starts with the initial specification of place and people \blackrectsmall{1}, then the AI carries on a conversation (\circledigit{a}). It segments what the adolescent has spoken (\circledigit{b}) in a specific structure (\circledigit{c}) and elicits missing parts via step-by-step questioning \blackrectsmall{2}. To aid autistic adolescents' comprehension and expression, the system provides a token-based visual representation of the current narrative in a four-panel comic strip \blackrectsmall{3}. (Please refer to our supplementary demo video at
\sourcecode{}.)}
  \Description{The figure presents an overview of AUTIVERSE and its three scaffolding layers for autistic adolescents' journaling with a peer-like AI. (1) Scaffolded settings: a setup screen asks the user to specify place and people to cue recall; an AI character and a teen illustration highlight a school incident (e.g., with "Oliver"). (2) Step-by-step questioning: a Q\&A interface shows short prompts that elicit missing details and segment the user's utterances. (3) Narrative + review: the central canvas displays a 2×2 four-panel comic that visualizes the story using the ABCE structure---Antecedent, Behavior, Consequence, Emotion—marked with color-coded labels along the left. Simple token-style figures represent characters and actions inside each panel. To the right of the comic, a conversation panel supports review and revision, offering prompts and action buttons (e.g., There's something to fix / It's fine). Dashed arrows and circled numerals depict the flow from initial specification → conversational elicitation → comic-based review.}
 \label{fig:teaser}
\end{teaserfigure}

\maketitle

\section{Introduction}

Adolescence is a critical developmental period for cultivating identity and independence~\cite{erikson1968identity}, characterized by increased autonomy from caregivers~\cite{steinberg1986vicissitudes}, deeper immersion in peer relationships~\cite{steinberg1986vicissitudes,brown2009peer}, and more sophisticated self-reflection~\cite{blakemore2014adolescence}. For autistic adolescents\footnote{In this work, we use identity-first language (\eg{}, autistic adolescent) rather than person-first language (\eg{}, adolescent with autism), considering the preferences of autistic individuals~\cite{Lorcan2016} and recent academic trends~\cite{Brian2002}.}, however, this period presents an additional set of cognitive, emotional, and communicative challenges~\cite{masoomi2025emotion}. Navigating heightened social risks such as peer conflict and bullying~\cite{sterzing2012bullying,donno2010social}, they often struggle to convey these experiences to their parents~\cite{kerr2000parents}. This challenge arises from the substantial cognitive demands of narrative construction---the process of translating complex daily events into a coherent verbal story~\cite{losh2003narrative,capps2000frog}.
\revised{Even HFA (High-Functioning Autism) adolescents with verbal abilities---those who can express their thoughts and understand others' speech---experience these difficulties, as the challenge lies not in basic language production, but in the cognitive process of selecting, sequencing, and structuring information for a listener~\cite{hill2004executive,happe2006weak}.}
Consequently, parents face a persistent dilemma: Fostering their child's increasing autonomy while simultaneously safeguarding them from harm~\cite{smetana1994adolescents}. Lacking reliable communication regarding daily events and associated social and emotional responses, caregivers often rely on fragmented or ambiguous self-reports, which complicate both understanding and timely intervention~\cite{norbury2003narrative}.

Journaling is a promising method to improve narrative skills, which can incrementally strengthen narrative competence by providing repeated opportunities to rehearse event organization, causal linking, and reflective meaning-making~\cite{danoff2010does, pennebaker1997writing}. However, it can be particularly burdensome for autistic adolescents due to its text-centric nature and high demand on executive functioning. Requiring autistic individuals---who often possess strong visual thinking skills~\cite{quill1997instructional, rao2006learning}---to express complex experiences exclusively through text overlooks their inherent cognitive strengths. In addition, the unstructured nature of the blank page imposes significant demands on executive functions, such as planning, sequencing, and language formulation~\cite{hill2004executive}. Therefore, autistic adolescents may benefit from alternative journaling approaches that provide structure, reduce linguistic demand, and incorporate visual aids that facilitate their storytelling, building on their strengths in visual thinking.

Yet, research on supporting autistic adolescents' journaling is largely underexplored, and only a handful of studies have explored the potential of structured self-tracking~\cite{Li2010StageBasedModel} of challenging behaviors (\eg, \cite{kim2019toward}) rather than freeform journaling of daily experience. 
Outside the autism context, the HCI community has explored scaffolding in journaling for children or adolescents, often leveraging Large Language Models (LLMs). These include chatbots that assist documenting daily experiences~\cite{kim2024mindfuldiary} or prompt children to share emotions~\cite{seo2024chacha}. However, they often rely on text-only dialogues as the primary interaction modality.

In this work, we investigate ways to facilitate autistic adolescents' journaling to help them organize facts and feelings surrounding their daily events. \revised{Here, we focus on autistic adolescents with functional verbal abilities---those who can speak but struggle to organize their experiences into coherent narratives due to executive functioning challenges rather than basic language difficulties.} Inspired by previous research demonstrating the benefits of using LLMs to provide conversational scaffolding across diverse topics~(\eg,~\cite{seo2024chacha, kim2024mindfuldiary}) as well as to organize structured information from freeform dialogues~(\eg,~\cite{choi2025aacesstalk, shin2025planfitting}), we explore how LLMs can support autistic adolescents in narrative construction.
To better understand their specific needs and preferences, we first conducted formative interviews with five autism experts and six parents of autistic adolescents. We found a strong need for structured scaffolding to guide narrative construction and mitigate the difficulty of responding to open-ended prompts (\eg{}, ``\textit{How was your day?}''). Participants emphasized the importance of visual aids in offloading the cognitive burden of verbal expression and expressed a desire for adolescents to engage in journaling autonomously, without excessive parent mediation that may hamper their sense of ownership.

Informed by these findings, we designed and developed \sysname{}, a tablet-based AI-guided multimodal journaling system (\autoref{fig:teaser}). \sysname{} leverages LLMs to generate conversational scaffolding, analyze information fragments, and generate comic strips in token-based visualization. 
\revised{By designing the chatbot to have a peer persona---a virtual character around the same age as the user---}\sysname{} engages autistic adolescents in a spoken/text conversation to gather information about an event of the day, incorporating the ABC (Antecedent-Behavior-Consequence) model~\cite{skinner1965science} along with emotion expression~\cite{gottman1998raising}. Based on the collected information, the system presents a story in the form of a 4-panel comic strip, which the adolescent reviews and revises in collaboration with the system. It serves as a visual aid that could potentially reduce reliance on linguistic expression while promoting intuitive, reflective storytelling.


To examine how \sysname{} supports autistic adolescents to recount their daily experience and its emotional impact, we conducted a two-week home deployment study with 10 autistic adolescent-parent dyads in South Korea. During the deployment period, adolescent participants journaled by conversing with an AI peer using \sysname{}, while a parent sat nearby in a safeguard role. Our results indicate that \sysname{} scaffolded coherent narratives and surfaced additional details about events and emotions. The AI peer provided a comfortable, nonjudgmental context for disclosure, making sharing feel enjoyable and under adolescents' own control. Parents, in turn, learned new information about their child's day and used it as a conversational bridge at home. These findings suggest a practical, family-compatible pathway for AI-based journaling that can support narrative development and reflection in everyday life.
The contributions of this paper are fourfold:
\begin{enumerate}[leftmargin=*, itemsep=4pt, topsep=2pt]
    \item Empirical insights from a formative study with autism experts and parents, identifying key challenges of eliciting narratives from autistic adolescents and design considerations for journaling systems in the field.
    
    \item The design and implementation of \sysname{}, a novel AI-guided multimodal journaling system that incorporates a customizable AI peer that helps scaffold narrative construction for autistic adolescents through conversation with visual supports. The source code and a demo video of \sysname{} is publicly available at \sourcecode{}.

    \item Empirical findings from a two-week deployment with 10 autistic adolescent-parent dyads, including direct accounts from the adolescents, illustrating how \sysname{} scaffolded journaling and its perceived effects on adolescents and parents.

    \item \revised{Design implications for AI-scaffolded journaling systems that foster autistic adolescents' multimodal narrative practice, highlighting the needs for supporting adaptive scaffolding to accommodate a wide autism spectrum and for socio-emotionally appropriate AI peers.}
\end{enumerate}

\section{Related Work}
In this section, we first cover the literature on social communication challenges of autistic adolescents. We then discuss journaling as an established, though demanding, method for narrative construction. We conclude by reviewing prior work on both the use of visual aids to mitigate cognitive load and the role of conversational AI as a collaborative partner for journaling.

\subsection{Social Communication Challenges of Autistic Adolescents}
Autistic adolescents demonstrate differences in social communication and interaction due to fundamental variations in cognitive processing that cause persistent challenges~\cite{edition2013diagnostic,frith2003autism}. For example, they often struggle to share attention with another person toward an object or event (\ie{}, joint attention)~\cite{mundy1994theory}, tend to focus on details rather than comprehending broader contexts or constructing a cohesive narrative~\cite{happe2006weak,jarrold2000linking}. Additionally, deficits in Theory of Mind---the ability to infer and reason about others' mental states---can make it difficult to anticipate how one's words and actions will be interpreted by others~\cite{baron1985does}.

These underlying cognitive characteristics manifest in tangible challenges during everyday social exchanges. Autistic individuals often struggle to interpret subtle nonverbal cues, such as facial expressions or prosody, which typically convey emotional nuance or contextual meaning~\cite{harms2010facial}. In parallel, a high prevalence of alexithymia---difficulty identifying and articulating one's own emotions---further limits their expressive capacity~\cite{bird2013mixed}. As a result, interactions that seem routine to neurotypical peers, such as recounting the day's events, may become cognitively and emotionally demanding tasks for autistic adolescents. Consequently, the dynamic and reciprocal nature of conversation~(\eg{}, initiating exchanges, turn taking, and maintaining topic coherence) can be particularly strenuous~\cite{saulnier2024essentials}.

These challenges become especially salient during adolescence, a developmental stage where nuanced peer relationships and emotional self-disclosure are increasingly important~\cite{brown2009peer,vijayakumar2020self}. Communicative differences during this period can act as significant barriers to forming friendships, participating meaningfully in classroom interactions, and avoiding social isolation~\cite{bauminger2000loneliness, williams2007social}. One area where these difficulties are particularly pronounced is personal storytelling, which requires the speaker to recall, organize, and express past experiences in a coherent and emotionally intelligible manner. Prior work has shown that narratives produced by autistic individuals are often fragmented, overly literal, or lacking in emotional context~\cite{norbury2003narrative,capps2000frog}. This hinders their ability to share inner experiences with others, limiting opportunities for relational development, empathy-building, and social learning.

\subsection{Journaling for Narrative Development in Autistic Individuals}
Journaling is known to be a viable method to strengthen narrative skills through repeated practice with event structuring, causal linkage, and reflective meaning-making~\cite{danoff2010does, pennebaker1997writing}. This process of externalizing thoughts enables individuals to organize fragmented memories into a coherent personal narrative, which in turn fosters a stronger sense of identity by allowing them to derive deeper meaning from their life events~\cite{waters2015relations,pennebaker1999forming}.

From a developmental psychology perspective, these benefits are particularly salient for autistic adolescents. For them, journaling strengthens the autobiographical memory and narrative identity~\cite{losh2003narrative,harvey2025narrative,wantzen2021autobiographical} and facilitates the training of emotional expression~\cite{mazefsky2013emotion}. 
However, HCI research in this area remains sparse~\cite{samson2015emotion,hu2024object,putnam2020children}, often focusing on capturing structured data about predetermined phenomena (\eg{}, mom's nagging~\cite{kim2019toward}) rather than open-ended narratives~\cite{kim2019toward,jo2022genieauti}.
This gap is consequential because the unique characteristics of autistic adolescents pose barriers to practice journaling themselves; they often feel difficulty organizing their experience into chronologically ordered and causally coherent narratives, with prior work documenting challenges across coherence, evaluation, and pragmatic structure~\cite{losh2003narrative,diehl2006story}.

To address these challenges, we propose a journaling system that lowers these cognitive demands while supporting their autonomy, with the following two key components: (1) visual support for structuring experiences, and (2) a conversational AI that offers scaffolding while encouraging self-disclosure. We detail each in light of prior work in the following sections.

\subsection{Visual Aids to Reduce Cognitive Demands}
A core principle in autism intervention is that pairing information with congruent visual aids can significantly reduce structural and cognitive demands by leveraging strengths in visual processing~\cite{quill1997instructional,rao2006learning}. For example, canonical autism intervention tools such as Social Stories~\cite{gray2000new} and Comic Strip Conversations~\cite{gray1994comic} recommend using images that align with the content of the guidance and conversation, to engage the autistic individuals and enhance their comprehension. Similarly, many frameworks for communication with autistic individuals (\eg{}, SCERTS~\cite{prizant2003scerts} and TEACCH~\cite{mesibov2010teacch}) emphasize visual mediation to streamline communication, such as symbol cards, photographs, and visual diagrams; a classic example is a visual schedule, which uses a sequence of picture cards (\eg{}, `Breakfast,' `School Bus,' `Class') to make the day's events predictable and understandable in TEACCH~\cite{mesibov2010teacch}.
High-tech adaptations, such as animations for emotion recognition~\cite{golan2010enhancing} and tablet-based prompt apps~\cite{fletcher2016designing,allen2016ipads}, further increase engagement and facilitate targeted skill acquisition. However, a common thread unites these approaches: they are primarily designed for receptive communication---to teach skills or convey information to the autistic individual. They offer limited support for expressive communication, specifically for constructing and conveying their own narratives.

Inspired by the benefits of visual aids to supplement narrative construction, this work synthesizes journaling and visual aids.
In contrast to recent youth journaling tools that remain largely text-based~\cite{kim2024mindfuldiary,seo2024chacha}, our system incorporates the token-based visual representation of narrative in a four-panel comic strip format, which is co-constructed through conversation with \revised{a chatbot}, aiming to retain autonomy and ownership while easing cognitive demands in organizing and expressing daily experiences.

\subsection{Conversational AI as a Collaborative Partner}\label{rw:conversational_AI}
Conversational agents shift narrative support from static templates to mixed-initiative interaction, enabling the \revised{chatbot} to elicit details, repair omissions, and propose structure dynamically. In mental health and education domains, context-based dialogue systems have proven effective in scaffolding reflection and reasoning, such as CBT micro-interventions~\cite{fitzpatrick2017delivering} and mixed-initiative tutoring~\cite{fitzpatrick2017delivering}. Within HCI, recent youth-focused systems have begun to utilize large language models (LLMs) to support journaling and emotional disclosure, demonstrating that lightweight conversational prompts can reduce initiation costs and sustain engagement~\cite{kim2024mindfuldiary,seo2024chacha}.

\begin{revisedenv}
In autism research, chatbots have typically served one of two roles: either as \textit{instructors} for training social communication skills~\cite{kandalaft2013virtual,shamsuddin2012humanoid,wainer2014using,puglisi2022social,bernardini2014echoes,mower2011rachel,ali2020virtual}, or as \textit{elicitors} of personal narratives~\cite{tartaro2008playing,bernardini2014echoes,kumazaki2018can}. Despite their prevalence, AI systems in conventional instructor roles (\eg{}, teachers) may inherit drawbacks reported in prior work with human instructors: autistic adolescents may respond defensively to perceived instructional attempts from authority figures~\cite{pappagianopoulos2025therapy,esqueda2025teacher}, which can impede engagement and learning. Meanwhile, elicitor systems, though less threatening, remain confined to passive interviewing---prompting for content without actively supporting its organization~\cite{kumazaki2018can}. Consequently, autistic adolescents must structure their narratives on their own, which is particularly challenging given the executive function difficulties~\cite{hill2004executive,petersen2014systematic}.

Building upon these lessons in prior research, we designed a chatbot to exhibit a \textit{peer} persona (\ie{}, a virtual character of the same age) that operates as a \textit{collaborative partner}. Peer-mediated interventions have demonstrated superior outcomes by positioning support as interaction between equals rather than hierarchical instruction~\cite{chang2016systematic}. Adopting a non-hierarchical relational stance, the peer-like framing can mitigate defensiveness by fostering rapport, increasing self-disclosure, and reducing resistance to engagement~\cite{kumazaki2022android}. 
In addition, as a collaborative partner, the \revised{chatbot} can actively work with the user to organize and structure their expressed thoughts into a coherent narrative, rather than merely asking questions to collect discrete information.
\end{revisedenv}

\section{Formative Study}
To inform the design of \sysname{}, we conducted formative interviews with five autism experts and six parents of autistic adolescents. We aimed to (1) understand the challenges that autistic adolescents and their communication partners (\ie{}, parents, teachers, or therapists) face in engaging in everyday conversations and (2) identify the strategies used to foster more natural and meaningful exchanges. We tailored the interview protocols for each group---experts and parents---and conducted remote interviews separately to elicit their unique experiences and perspectives. We recruited participants from both groups through snowball sampling and via the internal network of one researcher, who is an autism specialist.

\subsection{Procedure and Analysis}

    \subsubsection{Interviews with Autism Experts}
    We recruited five autism experts (E1--5) who have years of experience and expertise in communicating with autistic adolescents. The experts included two special education professors, one special education teacher, one autism-focused music therapist, and one developmental assessment specialist from a psychiatry department. The experts had an average of 19 years of experience (ranging from 15 to 23 years), and all of them had clinical experience with autistic adolescents in promoting their language and behavioral development.
    
    We first asked the experts about the characteristics and challenges that autistic adolescents face in everyday conversations, as well as the strategies that experts use to understand their intentions and facilitate reciprocal interactions. We then presented a storyboard as a slideshow to help the experts understand the concept of an AI-guided multimodal journaling system. The storyboard prototype depicted a scenario in which a parent and an autistic adolescent are sitting in front of a tablet side by side, while the adolescent engages in a conversation with \revised{a chatbot with a peer persona,} in the system to create a draw-and-write journal about ``\textit{what happened today.}'' This scenario was developed in consultation with one of the authors, who is an expert in autism communication strategies with extensive counseling and clinical experience. We asked the experts about clinically appropriate AI behaviors, conversational directions, design considerations to support unique usability needs for autistic adolescents, features that could enhance engagement for repeated use, and minimize potential risks. The interviews lasted about 1 to 1.5 hours. We offered a 100,000 KRW (approx. 72 USD) gift card as compensation.

    \subsubsection{Interviews with Parents of Autistic Adolescents}
    \revised{We recruited six parents (M1--6; all mothers) of autistic adolescents who met the following inclusion criteria aligning with our target population: (1) a clinical diagnosis of Autism Spectrum Disorder (ASD) classified as Level 1 (formerly described as high-functioning autism), and (2) the ability to express their thoughts and understand others' speech without major difficulty, while still experiencing persistent challenges in sustaining everyday conversations. Five adolescents were boys and one was a girl, with an average age of 15.67 years (ranged 13--18, $SD=1.75$).}
    The semi-structured interview began with a discussion about the characteristics and challenges of having daily conversations with their child. We then asked them to share the efforts they had made to improve these conversations and the strategies that led to successful outcomes. As with the expert interviews, we presented the same storyboard as a probe to elicit their perspectives on design considerations from a parental standpoint, as well as the potential roles AI could play in eliciting everyday experiences from their child. Each interview lasted about an hour, and participants received a gift card worth 50,000 KRW (approx. 36 USD) as compensation.


    \subsubsection{Analysis}
    All interviews were audio-recorded and later transcribed. Applying thematic analysis~\cite{braun2006using}, one researcher open-coded the transcript to identify emerging themes. The entire research team then finalized the themes through multiple rounds of discussion. In the following, we present the findings from the formative study.

\subsection{Finding 1: Difficulty in Open-Ended Daily Conversations}
Both parents and experts reported that autistic adolescents struggle with open-ended daily conversations, especially when asked to initiate or structure their own narratives. Two interrelated difficulties emerged: (1) cognitive overload triggered by non-specific prompts, and (2) a subsequent challenge in presenting experiences coherently for a listener.

    \subsubsection{Struggling to Answer Non-Routinized Open-Ended Questions}
    Autistic adolescents suffered from cognitive overload induced by broad questions such as ``\textit{What happened today?}''. Such open-ended prompts place high demands on executive functioning, requiring adolescents to select and structure a single experience from countless possibilities. E5 remarked, ``\textit{When someone asks, `What did you do today?', they don't know what to choose to talk about among the thousands of things they did. [...] I ask short questions to help them decide what to say.}''

    This difficulty was especially noticeable when adolescents received new inquiries (E1, E3--4, and M4), while they could often respond to familiar prompts. E4 noted, ``\textit{When it comes to recalling something new or bringing up experiences they had on their own, that's really difficult for them.}'' This difficulty with these inquiries points to the importance of a structured format for these adolescents. As the participants observed, unfamiliar questions without structured cues can lead to communicative breakdowns, reinforcing their reliance on predictable conversational patterns. E3 stressed, ``\textit{Open-ended or reflective questions often lead to a sense of failure for them, so it's important to begin with things they can answer to help them feel more secure.}''


    \subsubsection{Struggling to Describe Daily Experiences}
    Both groups of participants consistently reported that autistic adolescents often struggled to communicate recalled events coherently. As a result, caregivers found it challenging to sequence events and identify relevant details, as they had to interpret and reconstruct meanings from incomplete and disjointed pieces of a narrative. M6 noted, ``\textit{What is most challenging is that I have to piece clues together and make inferences like `Ah, maybe he did this because of that.'}''

    This difficulty seems to reflect autistic adolescents' core challenge in organizing lived experiences into coherent and structured accounts (E1--5, M1--2, M4, and M6). 
    E1 explained, ``\textit{If I give them a clear plot, they can do really well. But if I just say `Tell me what happened,' they jump from one thing to another. They have trouble organizing their experiences into a coherent story.}'' This was often compounded by a highly self-referential perspective (E1, E4, M1, and M3), with adolescents showing ``\textit{a tendency to focus on themselves rather than attending to the surrounding context or engaging in reciprocal conversation}'' (E4).

\subsection{Finding 2: Visual Scaffolds for Eliciting Daily Narratives}\label{sec:f2}
All participants reported that visualization or visual materials (\eg{}, drawings, photos, or objects) were highly effective for eliciting detailed responses by making abstract memories more concrete and reducing cognitive load. As M4 noted, ``\textit{Giving even a simple visual cue, like drawing a comic, helps the child share his thoughts in much more detail. I realized how effective those visual prompts can be.}''

Although often effective, current visual methods were also constrained by various factors. For example, requiring adolescents to draw from scratch can be counterproductive, particularly for those with anxiety or perfectionistic tendencies. E1 explained, ``\textit{Some kids with co-occurring anxiety disorders hesitate to even begin, fearing that they might not draw well.}'' Participants also pointed out that existing visual materials are often too rigid or static to represent emotionally nuanced situations (M3--4). M3 commented, ``\textit{The available visual materials are quite limited, making it difficult to present them flexibly and appropriately in each situation.}''

Beyond these limitations, participants expressed differing views on the appropriate level of visual detail. While some (M1, M5) advocated for rich, `Studio Ghibli'-like illustrations for engagement (M1), the majority (E2, E4--5, M2, M4, M6), particularly experts, argued that simpler visual representations are generally more effective. E4 explained, ``\textit{Here, the symbols themselves are not the key---the important thing is that the child recalls their day. If the illustrations are too detailed, children may get lost in those details rather than focusing on their memories.}'' Similarly, M2 noted, ``\textit{If the visual is too detailed and differs from what the child imagined, it could lead to confusion. However, many children are already familiar with symbolic representations since they often encounter them in therapy settings.}''

\subsection{Finding 3: Balancing Parental Involvement while Cultivating Independence}

Some parents (M2--3) emphasized that the long-term goal of journaling should be fostering autistic adolescents' independent and self-initiated narration. In reality, as these adolescents often require co-regulation and safety monitoring~\cite{ting2017emotion,prizant2003scerts}, parents usually stay with them or nearby. As a result, it appeared to be challenging for parents to negotiate the balance between supporting independence and continuing parental involvement. M2 remarked: ``\textit{My child's level of dependency on me is unlike that of most adolescents. The goal is to support their independence by stepping back, so our relationship is quite different from that of typical parents and teens.}''

\revised{This tension revealed a broader pattern: autistic adolescents are predominantly surrounded by authority figures---parents, teachers, therapists---whose interactions tend to involve instruction or correction. Participants suggested that an AI positioned as a peer, rather than an authority, could create a fundamentally different relational context as equals. M6 explained the contrast vividly: \textit{Instead of someone trying to teach, AI should be like a real peer. My child really listens to friends---they joke around and share things. A truly peer-like AI would be wonderful.}'' E5 reinforced this point: ``\textit{Teachers always have rules about what to do or not do, but a same-age peer can talk comfortably, and the child might open up more.}''}

Regarding the idea of AI-guided journaling, participants emphasized that parents should play a subsidiary role, as the ``\textit{ultimate goal is to do it on [child]'s own}'' (M5). Experts further stressed that in journaling contexts---where adolescents articulate their experiences and thoughts---parental involvement should be carefully designed to minimize proactive involvement. Without clear guidance, they warned, parents may ``\textit{unintentionally engage in negative forms of involvement}'' (E3)---a risk amplified in autism, where adolescents show increased sensitivity to negative parental attitudes~\cite{baker2019parental}. To mitigate this, experts recommended that systems explicitly encourage parents to ``\textit{observe, instead of participate,}'' as excessive intervention could lead adolescents to resist using the tool (E4).


\section{\sysname{}}
Informed by the formative interviews, we designed and developed \sysname{}, an AI-guided multimodal journaling system for autistic adolescents. In this section, we describe the design rationales derived from the formative interviews, present the user interface and system components of \sysname{}. We then illustrate the generative pipelines in the system, along with implementation details.

\subsection{Design Rationales}
    \ipstart{DR 1. Structure Experience through Routines}\phantomsection{}\label{sec:dr1}
    Building on formative insights and prior works, we designed a structured narrative scaffold---instead of broad, open-ended prompts---to help adolescents organize daily experiences. We initiate conversations with concrete situational anchors (\eg{}, location, people involved) to reduce ambiguity and support memory retrieval, in line with evidence on supportive questioning for autistic communicators~\cite{norris2022supporting}. To support narrative construction, we designed the system to follow a stepwise dialogue structure inspired by \textit{ABC model}~\cite{skinner1965science}, an evidence-based approach that frames behavior by breaking it into \textbf{antecedent} (environmental context), \textbf{behavior} (the main event), and \textbf{consequence} (the result of the event). Given the importance of emotional awareness of both neurodiverse~\cite{huggins2021emotional} and neurotypical~\cite{seo2024chacha, gottman1998raising} children, we appended \textbf{emotion} to the three stages to foster emotional awareness and reflection. (We refer to this extended model as the ABC-E format hereinafter.) 
    By internalizing this scaffold structure, we aimed to encourage adolescents to develop narrative competence by learning how to organize and share their experiences in socially intelligible ways.

    \ipstart{DR 2. Balance Flexibility and Executive Load through Visual and Conversational Scaffolds}\phantomsection{}\label{sec:dr2}
    Our formative interviews and existing literature~\cite{mattison2018drawing,van2011anxiety,hill2004executive} highlight the need for simplified visual scaffolds that maintain expressive flexibility while minimizing the executive load to produce them. To this end, we incorporated an AI-generated \textbf{four-panel comic strip} for journaling, with each panel corresponding to each component of the ABC-E format (see \hyperref[sec:dr1]{DR1}). In each panel scene, we used simplified token-based visual representations to describe it (see \blackrectsmall{3} in \autoref{fig:teaser}) to avoid unintended distractions caused by detailed images~\cite{sasson2008children}.

    We complement this visual support with a conversational scaffold, where a \revised{chatbot} guides the overall journaling process. The \revised{chatbot} uses \textbf{step-by-step questioning} to incrementally elicit the A, B, C, and E components through small, sequential prompts. This method reduces the cognitive demands of spontaneous storytelling by breaking the task into manageable units---a technique universally endorsed by our expert participants (E1--4) and aligned with prior AI-driven prompting methods~\cite{lee2023dapie,seo2024chacha}.  
    This dual-scaffold approach, combining the token-based four-panel comic strip and step-by-step conversational AI, preserves expressive flexibility while offloading executive and motor demands, enabling low-friction journaling for autistic adolescents.

    \ipstart{DR 3. Prioritize Adolescent-AI Interaction}\phantomsection{}\label{sec:dr3}
    Considering the tension between parental support and adolescent autonomy, we prioritized the adolescents' independence and designed the interface on the assumption that they would interact with the system on their own. We intentionally positioned the parent as a supportive observer to ensure safety while respecting their child's autonomy.
    \revised{We also designed the AI to have a peer persona---a virtual character around the same age as the user---that establishes a different relational dynamic from parental interaction. Drawing on the formative study findings and evidence that adolescents tend to be more receptive to informal, peer-style interactions than to adult-led instruction, particularly in emotionally sensitive contexts~\cite{chang2016systematic}, we positioned the AI as an equal rather than an authority. This creates a non-hierarchical space where adolescents can practice narrative construction without evaluation or social consequences, reducing the defensiveness often triggered by parental or teacher interactions and encouraging authentic self-expression~\cite{kumazaki2018can,tokic2011parental}. The peer persona is operationalized through customizable features---adolescents select the AI peer's name, voice, and appearance, which is known to enhance their reception and engagement~\cite{kao2022audio,kumazaki2017pilot}---and through conversational tone that communicates playfully and offers feedback in a friendly manner while actively helping organize the adolescent's narrative as a collaborative partner.}

\subsection{User Interface and Interaction with \sysname{}}\label{scenario}
\begin{figure*}[t!]
    \centering
    \includegraphics[width=0.97\textwidth]{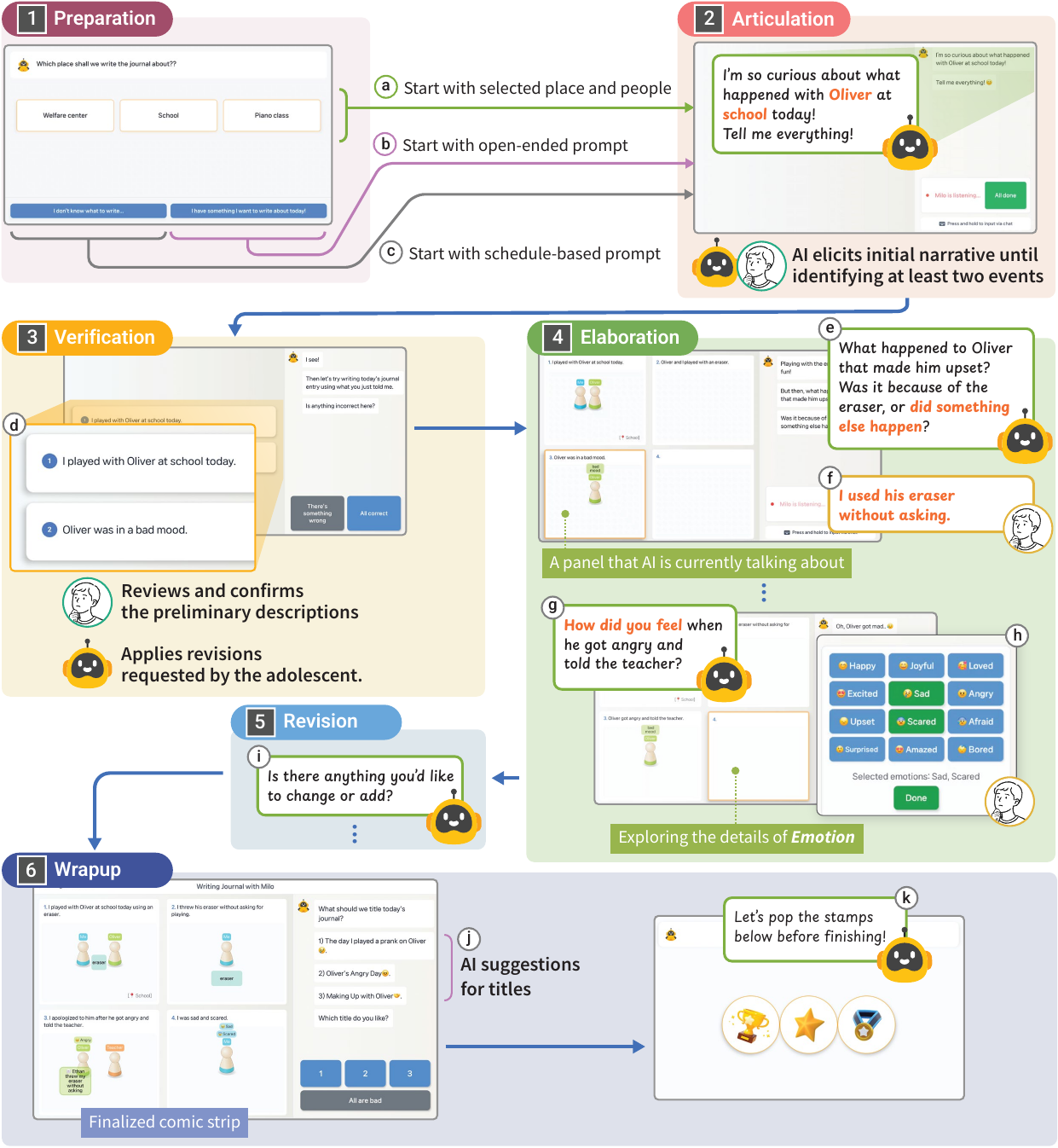}
    
    \caption{The usage flow of \sysname{}. \blackrectsmall{1} The user can begin journaling in one of the three starting modes: starting with a selected place and people (\circledigit{a}), an open-ended prompt (\circledigit{b}), or a schedule-based prompt (\circledigit{c}). \blackrectsmall{2} The AI elicits an initial narrative from the user. \blackrectsmall{3} The AI presents preliminary descriptions (\circledigit{d}) based on the user's narrative, which the user reviews and revises on demand. \blackrectsmall{4} The AI asks questions about missing details. If the user struggles to describe emotions, the AI provides a list of 12 example emotions (\circledigit{h}). \blackrectsmall{5} Based on the updated comic strips, the AI asks the user whether to make any final changes or additions. \blackrectsmall{6} To wrap up, the AI offers an empathetic response and title suggestions (\circledigit{j}) for the finalized narrative. Finally, the AI praises the user by awarding stamps (\circledigit{k}).
    }
    \Description{This figure is a flowchart illustrating the six-stage process for a user creating a journal entry within the AUTIVERSE application, with the flow moving from top to bottom and each stage numbered and accompanied by screenshots of the app's user interface and annotations explaining the interactions.
    - Phase 1: Preparation shows the user beginning the journal by choosing one of three prompt types: starting with (a) a selected place and people, (b) an open-ended prompt, or (c) a schedule-based prompt, each represented by a screenshot of the initial input screen.
    - Phase 2: Articulation depicts an AI character prompting the user to describe their day with the phrase "I'm so curious about what happened with Oliver at school today. Tell me everything!" The accompanying text explains that the AI elicits an initial narrative from the user until at least two events are identified.
    - Phase 3: Verification illustrates the AI presenting preliminary descriptions based on the user's narrative in a chat interface, showing examples like "I played with Oliver at school today" and "Oliver was in a bad mood," with text indicating that the user reviews and confirms these descriptions and can apply revisions requested by the user.
    - Phase 4: Elaboration shows the AI asking questions about missing details and emotions, with examples such as "What happened to Oliver that made him upset? Was it because of his temper, or did something else happen?" and "How did you feel when he got angry and told the teacher?" The AI provides a list of 12 emotion buttons for the user to select from, and the overall section is labeled "Exploring the details of Emotion."
    - Phase 5: Revision depicts the AI asking the user, "Is there anything you'd like to change or add?" as the final narrative takes shape, indicating the point where the user can make further modifications.
    - Phase 6: Wrapup concludes the process, showing the AI presenting the finalized narrative in a comic strip format, along with AI suggestions for titles, an option to "Put a title to today's journal," and a final interaction where the AI praises the adolescent with "Let's pop the stamps below before finishing!" and a visual representation of awarding stamps.}
    \label{fig:interface}
\end{figure*}

\autoref{fig:interface} illustrates the user flow of journaling with \sysname{}, where an autistic adolescent records a journal entry for the day. This process consists of six sequential phases: (1) \phasepreparation{}, (2) \phasearticulation{}, (3) \phaseverification{}, (4) \phaseelaboration{}, (5) \phaserevision{}, and (6) \phasewrapup{}. Each phase was carefully designed to complete a four-panel comic strip, grounded in the ABC-E format. By using a stepwise dialogue structure, we support users in recalling and organizing their daily experiences and emotions. In the following, we describe each phase and associated interactions by an imaginary autistic adolescent, Ethan: \textit{Every evening, 12-year-old Ethan sits at the desk and launches the \sysname{} app on a tablet to record his day through a four-panel comic strip. On the home screen, Ethan presses the \selectionbox{Start} button to start journaling for the day. His AI peer, Milo, appears and greets him.}

\ipstart{\phasepreparation{}: Identifying settings and characters (\blackrectsmall{1} in \autoref{fig:interface})} 
Milo first asks, ``\texttt{Which place shall we write the journal about?}'' along with a set of familiar locations Ethan often visits. Ethan selects \selectionbox{School}, recalling something that happened there. Milo then follows up with a question about who was involved, presenting a list of familiar people from school. Ethan chooses his friend, \selectionbox{Oliver}.

\ipstart{\phasearticulation{}: Collecting initial narrative (\blackrectsmall{2} in \autoref{fig:interface})}
Based on the selected place and people (\circledigit{a} in \autoref{fig:interface}), Milo responds playfully, ``\texttt{I'm so curious about what happened with Oliver at school today! Tell me everything!}'' Ethan replies, ``\textit{I played with Oliver. He was upset.}''

\ipstart{\phaseverification{}: Verifying preliminary structured narrative (\blackrectsmall{3} in \autoref{fig:interface})}
Milo processes the input and displays a structured outline on the left side of the screen: [\textbf{1}: \textit{I played with Oliver at school today}, and \textbf{2}: \textit{Oliver was in a bad mood}] (\circledigit{d} in \autoref{fig:interface}). Milo checks with Ethan to ensure shared understanding, asking: ``\texttt{I see! Then let's try writing today's journal entry using what you just told me. Is anything incorrect here?}'' Ethan presses the \selectionbox{All correct} button. Milo cheerfully continues, ``\texttt{Great! I'll draw your comic strip based on what you told me.}''

\ipstart{\phaseelaboration{}: Supplementing missing details in the ABC-E format (\blackrectsmall{4} in \autoref{fig:interface})}
A four-panel comic strip is generated and appears on the screen. There are two empty panels since the information corresponding to B and E was missing. Milo notes, ``\texttt{I couldn't draw everything with the information I had. Could you help me fill in the missing parts?}'', and Ethan agrees. After confirming that Ethan and Oliver played with an eraser, Milo prompts, ``\texttt{Playing with the eraser sounds like fun! But then, what happened to Oliver that made him feel bad? Was it because of the eraser, or did something else happen?}'' (\circledigit{e} in \autoref{fig:interface}). Ethan replied ``\textit{I used his eraser without asking.}'' (\circledigit{f} in \autoref{fig:interface}), and the panel B's description is updated accordingly. Milo then continues, ``\texttt{How did you use the Oliver eraser? Did you just erase it quickly, or did you do something else too?}'' Ethan replies, ``\textit{I threw it and played.}'' 
Milo follows up, ``\texttt{Oh, I see. What was Oliver's reaction when he saw you throwing the eraser? Did he get angry, or did he say something?}'' Ethan answers, ``\textit{He got angry and told the teacher.}'' The panel description is updated again. Milo now asks about the emotion to elicit a clue for panel E: ``\texttt{Oh, Oliver got mad.. How did you feel when he got angry and told the teacher?}'' (\circledigit{g} in \autoref{fig:interface}). A set of 12 emotion cards appears on the screen (\circledigit{h} in \autoref{fig:interface}). Ethan looks at the cards and selects \selectionbox{sad} and \selectionbox{scared} buttons.

Milo responds warmly, ``\texttt{Thanks for answering my questions. Now I can draw in the rest of the panel!}'' The panel descriptions are updated as follows: [\textbf{A}: \textit{I played with Oliver at school today using an eraser}, \textbf{B}: \textit{I threw his eraser without asking for playing}, \textbf{C}: \textit{Oliver got angry and told the teacher}, and \textbf{E}: \textit{I was sad and scared}].

\ipstart{\phaserevision{}: Reviewing and refining complete journal (\blackrectsmall{5} in \autoref{fig:interface})}
With the updated four-panel comic strip, Milo asks, ``\texttt{Is there anything you'd like to change or add?}'' (\circledigit{i} in \autoref{fig:interface}). Ethan presses \selectionbox{There's something to fix} button. Milo asks ``\texttt{Which part do you want to change, and how?}'' and Ethan replies, ``\textit{I apologized to him.}'' With this addition, Milo immediately updates the panel C. ``\texttt{Is everything correct now?}'' Milo asks. Ethan confirms and presses the \selectionbox{Yes} button.

\ipstart{\phasewrapup{}: Putting a title and rewarding the adolescent (\blackrectsmall{6} in \autoref{fig:interface})}
Milo thanks Ethan for sharing his story and responds based on the completed panel description, ``\texttt{So you and Oliver had a falling out at school and hurt each other's feelings. It breaks my heart to hear you felt sad and scared. I'll be rooting for you to have better things happen next time!}'' Then Milo says ``\texttt{What should we title today's journal?}'' listing three candidate titles (\circledigit{j} in \autoref{fig:interface}). Ethan chooses the first one, `\texttt{The day I played a prank on Oliver.}' Milo says ``\texttt{I'm glad you liked `The day I played a prank on Oliver'! Once you've checked the completed journal on the left, let's move on to the next one!}'' After reviewing the finalized comic strip, Ethan presses the \selectionbox{Next} button. Then, Milo awards Ethan three stamps as a reward for finalizing a session (\circledigit{k} in \autoref{fig:interface}).

\subsection{Conversational Pipelines}\label{conv_pipelines}
\begin{revisedenv}
Here, we describe the conversational pipeline and the model prompting strategies to support safe interaction with autistic adolescents.

\subsubsection{Pipeline Architecture}
Each phase of journaling, except the \phasepreparation{} phase, involves a conversational loop carried by an LLM, with additional LLM-infused components for dialogue analyses and structured information generation. \autoref{fig:system:pipeline:simplified} illustrates the flow of conversational pipeline.
In the \phasearticulation{} phase, the chatbot aims to identify at least two pieces of events from the adolescent's narratives. The adolescent is then allowed to modify the preliminary event descriptions (\circledigit{a} in \autoref{fig:system:pipeline:simplified}) in the next \phaseverification{} phase. The system organizes the resulting descriptions into an interim comic script (\circledigit{b} in \autoref{fig:system:pipeline:simplified}), which may contain incomplete or misplaced panels according to the ABC-E format. In the \phaseelaboration{} phase, the chatbot tries to complete the comic strip by collecting more information from the adolescent or asking for clarification. The chatbot messages are generated based on the inspection of an LLM-driven \textbf{Story Analyzer} (\circledigit{c} in \autoref{fig:system:pipeline:simplified}) that identifies issues regarding information or composition in the current comic strip. After receiving the adolescent's response, the \textbf{Description Reconstructor} (\circledigit{e} in \autoref{fig:system:pipeline:simplified}) updates the panel descriptions accordingly. Once the Story Analyzer assesses the current comic strip to be complete, the chatbot guides the adolescents to review and revise the complete comic strip (\circledigit{f} in \autoref{fig:system:pipeline:simplified}) in the \phaserevision{} phase. 
Lastly, in the \phasewrapup{} phase, the chatbot (1) presents a warm and personalized response based on the contents of the comic strip; and (2) proposes three candidate titles designed to be emotionally resonant and contextually appropriate, one of which to be chosen to conclude the journaling process. Refer to \autoref{appendix:pipeline} for technical details of the conversational pipeline.

\begin{figure*}[t]
    \centering
    \includegraphics[width=\textwidth]{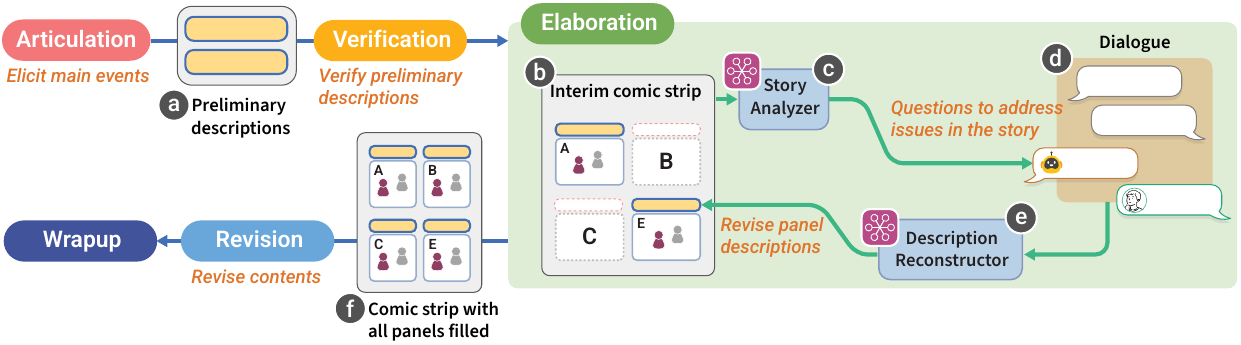}
    \caption{\revised{Conversational pipeline of a journaling session of \sysname{} after \phasepreparation{}. All phases involve LLM-driven conversational loop with specific chatbot goals. Refer to \autoref{appendix:pipeline} for more technical details.}}
    \Description{The figure illustrates the conversational pipeline of a journaling session in AUTIVERSE, depicting a left-to-right flow through five sequential phases after Preparation. Each phase is represented by a colored rounded-rectangle label: Articulation (coral/red, labeled "Elicit main events"), Verification (yellow/orange, labeled "Verify preliminary descriptions"), Elaboration (green header spanning a large shaded region), Revision (orange, labeled "Revise contents"), and Wrapup (dark blue). Between Articulation and Verification, element (a) shows two stacked document icons representing "Preliminary descriptions." The Elaboration phase occupies the largest area and contains an internal loop: (b) an "Interim comic strip" displayed as a 2×2 grid with panels labeled A, B, C, and E—panel B is highlighted to indicate current focus; (c) a "Story Analyzer" module (gear icon) that generates "Questions to address issues in the story"; (d) a "Dialogue" section showing speech-bubble icons for user–AI conversation; and (e) a "Description Reconstructor" module that outputs "Revise panel descriptions," cycling back to update the comic strip. Green arrows trace this iterative loop within Elaboration. After Elaboration, element (f) shows a fully completed comic strip with all four panels filled, feeding into Revision and finally Wrapup. Blue and gray arrows connect all major phases, illustrating the end-to-end LLM-driven conversational flow.}
    \label{fig:system:pipeline:simplified}
\end{figure*}

\subsubsection{Model Prompting Strategies for Safe Conversation}

Considering the characteristics of autistic adolescents, we carefully designed instruction prompts for LLM-generated responses in the \phasearticulation{} and \phaseelaboration{} phases (\cf{}, \circledigit{d} and \circledigit{i} in \autoref{fig:pipeline:description} in \autoref{appendix:pipeline}). Unlike other phases---\phaseverification{}, \phaserevision{}, and \phasewrapup{}---where the chatbot uses preset utterances to follow a predefined rule-based scenario, the two phases drive the conversation primarily through open-ended questioning to support spontaneous scaffolding. Our foremost ethical consideration in designing for these phases was to ensure that adolescents would not feel violated or overwhelmed by the AI's inquiries.

To this end, we first established basic conversational guardrails upon prior work on conversational AI for children~\cite{seo2024chacha} to prevent the chatbot from generating responses that could confuse or burden the adolescents. The chatbot was instructed to keep responses short (one to two sentences) and ask only one question per turn, as longer and more complex outputs increase the likelihood of the model generating irrelevant or erroneous content~\cite{zheng2025lvlms}. If adolescents ask questions beyond the scope of journaling (\eg{}, inquiring the bot of sensitive topics), we prompted the chatbot to avoid direct answers (\eg{}, ``\texttt{I don't know}'') and return to the conversation topic, rather than risking inappropriate advice or fabricated answers.

We then incorporated additional prompting strategies informed by formative interviews with domain experts, as well as consultation with the autism expert author, to address the specific needs of autistic adolescents.
First, we instructed the chatbot to use clear and literal language, avoiding both vague expressions (\eg{}, `things like that,' `at that time') that may induce anxiety through ambiguity~\cite{TagerFlusberg1999impairments} and figurative language such as metaphors and symbols that autistic adolescents often struggle to interpret~\cite{Happe1993figurative,happe2006weak}. In this way, we prevent confusion or misunderstanding that could make them feel overwhelmed.
Second, since direct causal `why' questions may trigger emotional dysregulation in autistic individuals~\cite{prizant2003scerts}, we instructed the chatbot to first follow up with details if an adolescent reports a socially problematic situation. The bot is to gather contextual information---such as what the individuals were doing---and environmental cues, including who was present nearby, before asking about causes or reasons. By beginning with concrete, straightforward questions, we aimed to help autistic adolescents prepare cognitively and emotionally for more explanatory questions.
Third, because autistic individuals may perseverate on topics of intense interest~\cite{anthony2013interests}, extended follow-up could derail the journaling process. To prevent this while maintaining rapport, we designed the chatbot to briefly acknowledge and redirect the conversation to the main narrative when adolescents introduce off-topic content such as special interests. We pre-configured each adolescent user's special interests (\cf{}, adolescent profile \circledigit{e} in \autoref{fig:pipeline:description}, Appendix \autoref{gp:main_event}) so that the chatbot can recognize these mentions without elaboration. 
Lastly, we limited the AI's self-disclosure and personality details. Although the chatbot adopts a peer persona, we intentionally designed it not to disclose personal stories, opinions, or backstory; it functions as a scaffolding partner for the adolescent's narrative rather than as a social companion seeking a mutual relationship. This design choice aimed to reduce the risk of emotional attachment while maintaining the peer-like rapport necessary for engagement.

\end{revisedenv}

\subsection{Implementation}
We implemented the core system of \sysname{} in Python using a FastAPI~\cite{fastapi_2025} server that serves REST APIs. The panel progress and conversation logs are stored in the PostgreSQL database. The generative pipelines incorporate OpenAI~\cite{openai_api_2025}'s ChatCompletion APIs to run the underlying LLM inferences. We used the \sloppy{\texttt{gpt-4.1-mini-2025-04-14}} model for generative tasks. \revised{To protect personal information included in model inputs, we used OpenAI Enterprise\footnote{https://openai.com/enterprise-privacy/}, which neither uses our input data for training nor retains it.} 

The client tablet app was implemented in TypeScript~\cite{typescript_2025} on React Native~\cite{reactnative_2025} as a cross-platform app running on both iPad and Android tablets. The client communicates with the server via REST API. On user turns, the utterance is audio-recorded and sent to the server. The recording is then transcribed using CLOVA Speech Recognition API\footnote{\url{https://api.ncloud-docs.com/docs/en/ai-naver-clovaspeechrecognition}}, which is widely applied to Korean automatic speech recognition. On AI turns, we used CLOVA Voice API\footnote{\url{https://api.ncloud-docs.com/docs/en/ai-naver-clovavoice}} to generate AI's voice for the corresponding utterance.
\section{Deployment Study}
We conducted a two-week field deployment study with 10 dyads of autistic adolescents and their parents. We aimed to examine how \sysname{} supports autistic adolescents to organize their daily experience and how the use of \sysname{} influences both adolescents and parents in perceptions and conversation patterns.
Here, we involved parents in two primary roles. First, to ensure safety while respecting adolescents' autonomy, we positioned parents as supportive observers, as mentioned in \hyperref[sec:dr3]{DR3}. Second, to address the challenges that autistic individuals can face with self-reporting and articulating their experiences in detail~\cite{keith2019importance,choi2025aacesstalk}, we engaged parents as primary informants. This methodology allowed us to gather rich, interpretative data from their recollections of their child's experience, which served as our main data source, complemented by the direct feedback adolescents provided in the exit survey.
The study protocol and materials were approved by the Institutional Review Board (IRB).

\subsection{Participants}

\begin{table}[b]
\sffamily\small
\def\arraystretch{1.2}\setlength{\tabcolsep}{0.15em}
\centering

\caption{Demographics of our participants in the deployment study, \revised{adolescents' IQ, and their communication characteristics (Communication) reported by} parents based on ~\cite{edwards2019adaptive}.}
\Description{This table summarizes parent–adolescent dyads in the deployment study and the adolescents' communication characteristics. The left block lists Parents (Alias P1–P10) and their Age; the middle block lists Autistic Adolescents (Alias C1–C10), Age (with gender), School (elementary, middle, or high), and IQ (values present for most rows and blank where unavailable). The next block contains nine columns labeled Communication characteristics 1–9; check marks indicate the characteristic is reported as present, and blanks indicate not reported; alternating light-gray column bands aid readability and do not encode data. Each row corresponds to one dyad (e.g., P1–C1 through P10–C10) and compiles all information for that pair. No personally identifying details are shown.}
\label{tab:participants}
\begin{tabular}{%
|c!{\color{lightgray}\vrule}%
c!{\color{gray}\vrule}%
c!{\color{lightgray}\vrule}%
c!{\color{tablegrayline}\vrule}%
c!{\color{tablegrayline}\vrule}%
c!{\color{tablegrayline}\vrule}%
*{9}{>{\centering\arraybackslash}m{0.015\textwidth}}|}
\hline

\rowcolor{lightgray}
\multicolumn{2}{|l!{\color{gray}\vrule}}{\textbf{Parents}} &
\multicolumn{13}{l|}{\textbf{Autistic Adolescents}}\\
\arrayrulecolor{lightgray}\hline

\rowcolor{tableheader}
& & & & & &
\multicolumn{9}{c!{\color{black}\vrule}}{\textbf{Communication}*}
\\

\rowcolor{tableheader}
\multirow{-2}{*}{\cellcolor{tableheader}\textbf{Alias}} &
\multirow{-2}{*}{\cellcolor{tableheader}\textbf{Age}} &
\multirow{-2}{*}{\cellcolor{tableheader}\textbf{Alias}} &
\multirow{-2}{*}{\cellcolor{tableheader}\parbox{1.2cm}{\centering{\textbf{Age}\\\textbf{(Gender)}}}} &
\multirow{-2}{*}{\cellcolor{tableheader}\textbf{School}} &
\multirow{-2}{*}{\cellcolor{tableheader}\textbf{IQ}} &
\cellcolor{tableheader}1 & \cellcolor{tableheader}2 & \cellcolor{tableheader}3 &
\cellcolor{tableheader}4 & \cellcolor{tableheader}5 & \cellcolor{tableheader}6 &
\cellcolor{tableheader}7 & \cellcolor{tableheader}8 & \cellcolor{tableheader}9 \\

\arrayrulecolor{black}\hline

\parent{1}    & 45             & \child{1}    & 12 (Boy) & Elementary                           & 81                         & & & & & & & &  \cellcolor{tablegrayline} \cellcolor{tablegrayline}\checkmark{}   & 
\\ \arrayrulecolor{tablegrayline}\hline
\parent{2}    & 46             & \child{2}    & 12 (Girl) & Elementary                          & 78      & \cellcolor{tablegrayline}\checkmark{} & \cellcolor{tablegrayline}\checkmark{} & & & \cellcolor{tablegrayline}\checkmark{} & & \cellcolor{tablegrayline}\checkmark{} & \cellcolor{tablegrayline}\checkmark{} & 
\\ \hline
\parent{3}    & 46              & \child{3}    & 11 (Girl) & Elementary                           & 70                         & & \cellcolor{tablegrayline} \cellcolor{tablegrayline}\checkmark{}  &  &  &  & \cellcolor{tablegrayline}\checkmark{} & \cellcolor{tablegrayline}\checkmark{} & & \cellcolor{tablegrayline}\checkmark{} 
\\ \hline
\parent{4}    & 50              & \child{4}    & 13 (Boy)   & Middle                         & 61      & &  &  &  & \cellcolor{tablegrayline} \cellcolor{tablegrayline}\checkmark{} & \cellcolor{tablegrayline}\checkmark{} & \cellcolor{tablegrayline}\checkmark{} & & \cellcolor{tablegrayline}\checkmark{}                                           
\\ \hline
\parent{5}    & 52              & \child{5}    & 13 (Boy)   & Middle                         & 63                         &
 & \cellcolor{tablegrayline}\checkmark{} & & & & & \cellcolor{tablegrayline}\checkmark{} & & \cellcolor{tablegrayline}\checkmark{} 
\\ \hline
\parent{6}    & 49              & \child{6}    & 14 (Boy)    & Middle                        & 71                         & 
 \cellcolor{tablegrayline}\checkmark{}  & \cellcolor{tablegrayline}\checkmark{} & \cellcolor{tablegrayline}\checkmark{} & & & & \cellcolor{tablegrayline}\checkmark{} & & 
\\ \hline
\parent{7}    & 47              & \child{7}    & 13 (Boy)   & Middle                         & 67                         &
 & \cellcolor{tablegrayline}\checkmark{} & & \cellcolor{tablegrayline}\checkmark{} & & & \cellcolor{tablegrayline}\checkmark{} & & \cellcolor{tablegrayline}\checkmark{} 
 \\ \hline
\parent{8}    & 54              & \child{8}    & 15 (Boy)   & Middle                         & 100      & \cellcolor{tablegrayline}\checkmark{} & \cellcolor{tablegrayline}\checkmark{} & \cellcolor{tablegrayline}\checkmark{} & \cellcolor{tablegrayline}\checkmark{} & & \cellcolor{tablegrayline}\checkmark{} & \cellcolor{tablegrayline}\checkmark{} & & 
\\ \hline
\parent{9}    & 49              & \child{9}    & 13 (Boy)   & Middle                         & 78                         & \cellcolor{tablegrayline}\checkmark{} & \cellcolor{tablegrayline}\checkmark{} & \cellcolor{tablegrayline}\checkmark{} & & & & \cellcolor{tablegrayline}\checkmark{} & &
\\ \hline
\parent{10}   & 56              & \child{10}   & 17 (Girl)   & High                        & 101                        & 
 & \cellcolor{tablegrayline}\checkmark{} & \cellcolor{tablegrayline}\checkmark{} & \cellcolor{tablegrayline}\checkmark{} & \cellcolor{tablegrayline}\checkmark{} & & & \cellcolor{tablegrayline}\checkmark{} &    
 \\ \arrayrulecolor{black}\hline                      
\end{tabular}%
\begin{flushleft}
\vspace{2mm}
\textbf{*Communication Characteristics~\cite{edwards2019adaptive}}
\end{flushleft}
\begin{enumerate}[leftmargin=*]
    \item Verbal and knows more words than just those used in their daily lives. 
    \item Have also learned vocabulary from other sources (\eg{}, reading, school, TV). \item More than just a functional vocabulary.
    \item Uses a variety of sentence types (simple to complex) and communicates opinions, ideas, news, events, aspirations.
    \item Might have significant difficulties in expressing ideas and feelings in words.
    \item Uses language to initiate and interact.
    \item Conversational difficulties might exist.
    \item Able to understand and use abstract language, but might have difficulty describing events in sequence.
    \item Can usually follow meaningful, simple, 3-step commands.
\end{enumerate}

\end{table}

We recruited autistic adolescent–parent dyads by advertising our study, distributing flyers to online communities of parents with autistic children, through a child development center in South Korea, where one of the authors is affiliated, and through snowball sampling. We established specific inclusion criteria to ensure the appropriateness and feasibility of participation. For adolescents, the criteria included: (1) a diagnosis of Autism Spectrum Disorder, classified as Level 1 (formerly referred to as high-functioning autism) under CDC guidelines, (2) the ability to express their thoughts and understand others' speech without major difficulty, albeit with persistent challenges in sustaining everyday conversations, and (3) no significant motor impairments, particularly in hand coordination, allowing for independent interaction with a touchscreen tablet. 
We restricted the participant age range to 10--17 years to align with early-middle adolescence (10--13 and 14--17) as defined in pediatric guidelines~\cite{AAP_Stages_Adolescence_2024}, and within the WHO definition of adolescence (10--19)~\cite{WHO_Adolescent_Health}. For families, we required (1) one parent to consistently accompany the adolescent during the two-week study as a supportive safeguard, and (2) a stable home Wi-Fi connection.

A total of 16 dyads expressed interest in participation. To minimize the risk of cognitive or emotional distress, we implemented two screening steps. First, we asked parents to submit written descriptions of their child's communication level and intelligence quotient (IQ). An autism expert reviewed these descriptions and IQ to assess system fit and flag potential concerns. Second, we shared a demo video illustrating \sysname{}'s interface and interaction flow and requested parent feedback regarding its usability for their child. After this process, two parents decided not to participate in the study. Furthermore, four dyads were excluded as they did not live in proximity to our institution, a requirement for the in-person delivery and setup of the study equipment. 

Ultimately, 10 dyads (\parentwrap{P1--10}, \childwrap{C1--10}; the same number indicates a parent–child pair) provided written informed consent and participated in the study, with no dropouts. While the initial screening questionnaires were primarily designed to target potential parent participants who aligned with IRB requirements, we also asked parents to confirm their child's willingness to participate. \autoref{tab:participants} summarizes participant demographics, including the adolescents' communication characteristics as reported by their parents. 
The adolescent participants' ages ranged from 11 to 17 years ($M = 13.3$); seven were male, and three were female. The parent participants' ages ranged from 45 to 56 ($M=49.4$); all were mothers who identified as the primary caregiver.
As compensation for their participation, we offered 250,000 KRW (approx. 182 USD) to each dyad.

\begin{figure*}[b]
    \centering
    \includegraphics[height=3.5cm]{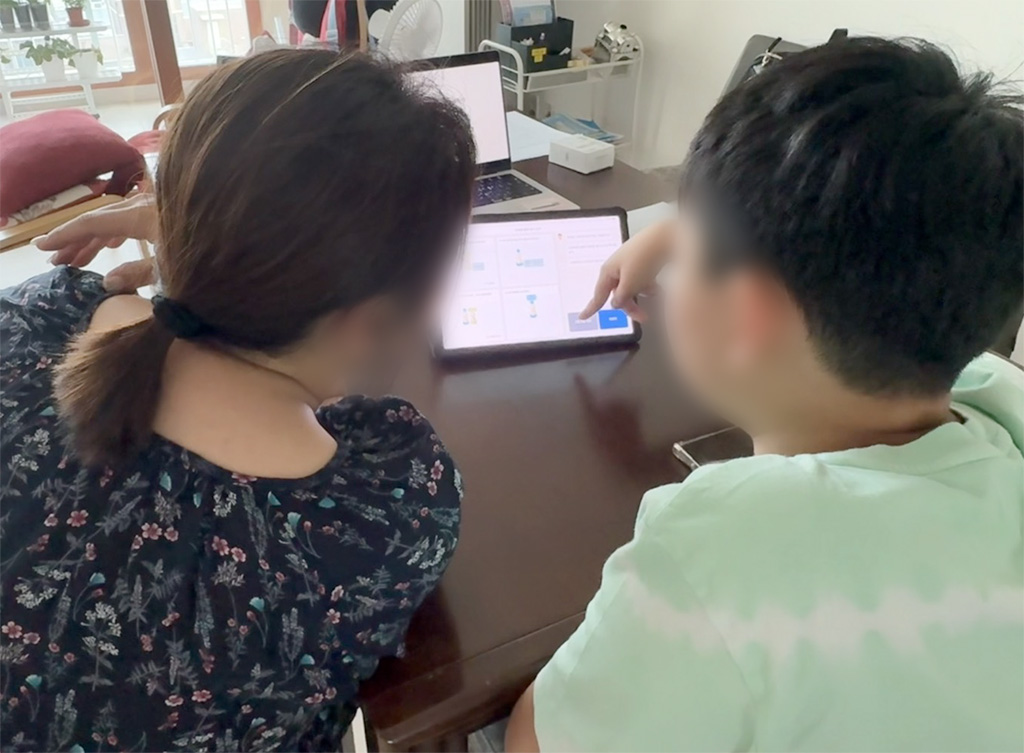}\hfill
    \includegraphics[height=3.5cm]{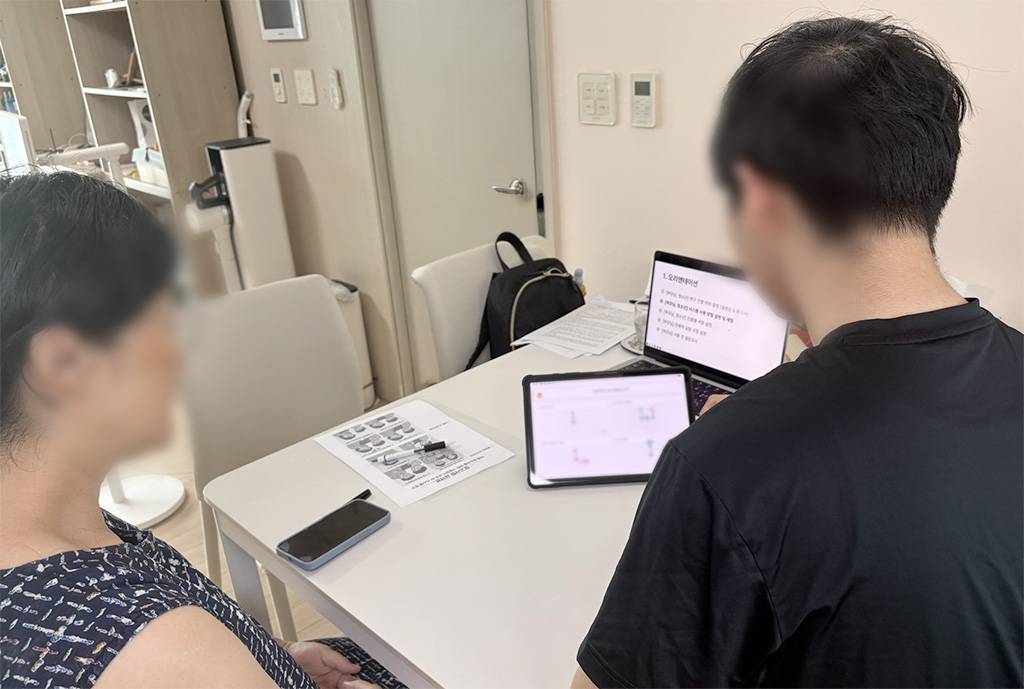}\hfill
    \includegraphics[height=3.5cm]{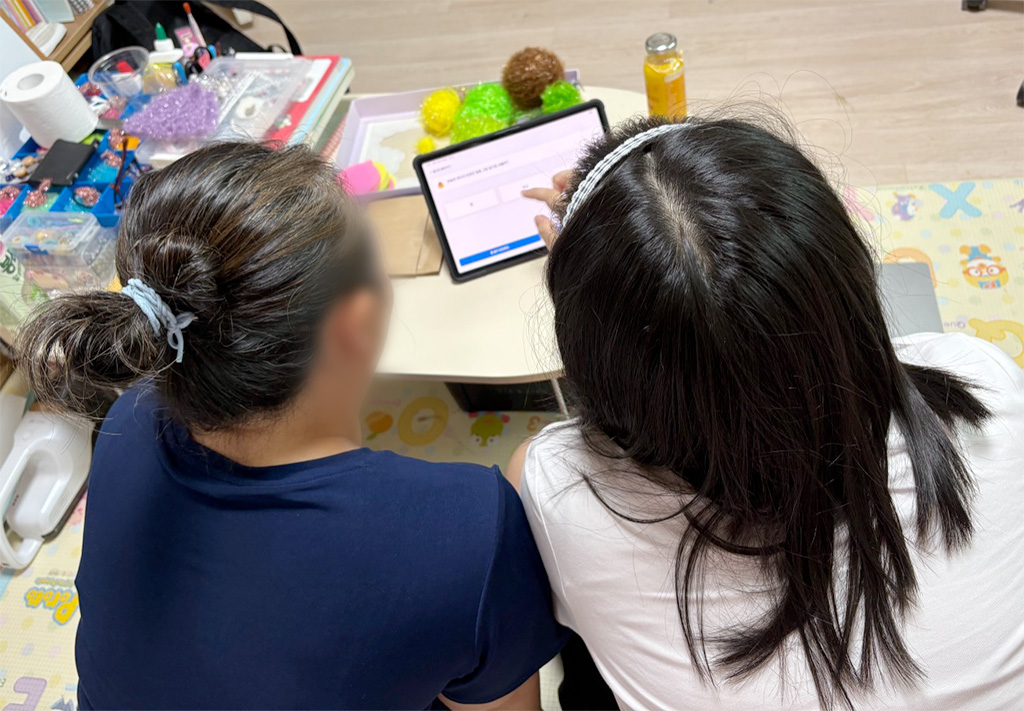}
    
    \caption{Parents and autistic adolescent participants are engaging in \sysname{} during the introductory session.}
    \Description{A triptych of candid photos shows parent–adolescent dyads using AUTIVERSE during an introductory session in home settings. Left: two participants sit side-by-side at a table, focusing on a tablet placed on a stand; a smartphone lies nearby. Center: another dyad sits at a dining table with a tablet and a laptop open; several sheets of paper are on the table. Right: two participants sit close together at a low table with a tablet in front; everyday items (e.g., stationery and small objects) are visible on the tabletop. All subjects face away from the camera and have blurred faces. Across images, the screens display the app interface, but specific on-screen text is not legible; the photos emphasize the co-located, side-by-side use of the system on consumer devices.}
    \label{fig:introductory_session}
\end{figure*}

\subsection{Procedure}
The study protocol consisted of four parts: (1) pre-study preparation, (2) introductory session, (3) two-week deployment of \sysname{}, and (4) debriefing.

    \ipstart{Pre-study Preparation}
    Before the study, we collected information through an online survey, asking parents to share their child's daily routines (\eg{}, locations visited and people encountered\revised{; see Appendix \ref{appendix:places}}) and personal interests to be entered in the system. We also asked participants to optionally customize the \revised{chatbot}'s name, voice, and visual representation. (We shared a link to CLOVA Voice\footnote{https://www.ncloud.com/product/aiService/clovaVoice?lang=en} where the dyads could listen to the sample of 25 available voices, and invited them to describe their preferred visual appearance---\eg{}, dinosaur, robot, doll, or realistic human.) These decisions were explicitly framed as the adolescent's choice to ensure alignment with their individual preferences.
    We then pre-configured the device and installed the \sysname{} app in advance to minimize the time for the initial setup. As a study device, we distributed a Samsung Galaxy Tab S9 tablet, featuring an 11-inch AMOLED display with a resolution of 1600 $\times$ 2560 (274 PPI). 

    \ipstart{Introductory Session}
    One researcher visited each dyad's home to configure the tablet device, connect it to the home Wi-Fi, and ensure that all necessary setup steps were complete (see \autoref{fig:introductory_session}). After explaining the goal of the study and the protocol, we obtained informed consent from both parents and adolescents. Then, both the parent and the autistic adolescent engaged in a pilot session, during which they freely created a sample journal entry using \sysname{} to support participants in understanding how to use the system.
    For parents, we provided specific guidelines for the deployment period: First, they were instructed to maintain a supportive but non-intrusive stance---allowing the adolescent to engage independently with the AI peer embedded in \sysname{}. 
    Second, they were asked to refrain from intervening in the conversation unless the adolescent explicitly requested assistance or asked a question.
    Lastly, parents were advised to adopt a positive and patient attitude and to avoid negative expressions that might discourage their child's engagement. 
    This guidance was designed to reinforce the adolescent's autonomy while preserving a sense of psychological safety. 
    The introductory session lasted approximately 40 minutes.
    
    \ipstart{Deployment} 
    From the day after the introductory session, participants began using \sysname{} at home over a two-week period. To allow participants to implement a natural and accessible daily routine, we allowed adolescents to engage with  \sysname{} at any time of day with no restrictions on the topic of choice. 
    We logged all interaction data, including the journal entries as well as the system's generated outputs and user responses.

    Once a day, we sent a text reminder to parents at their preferred time. In the evening, if a journal entry has been recorded for the day, we asked parents to complete a brief survey reviewing the daily experience. The survey asked parents to rate the level of parental moderation to support their child on a 5-point Likert scale. 
    The survey also asked whether \sysname{} helped parents learn new aspects of the adolescents' daily events or emotions, and whether they engaged in positive follow-up interactions (\eg{}, praise, additional questions) after journaling, all on 5-point Likert scales. Conditional open-ended items prompted elaboration on what the parents learned and how they assisted their child, if applicable.


    \ipstart{Debriefing}
    The day after the two-week deployment, we visited each household for an in-person debriefing session in which we administered surveys to adolescents and parents and conducted a semi-structured interview with parents. For adolescents, the exit survey comprised 11 items on a 5-point Likert scale: five adapted from the Technology Acceptance Model (TAM)~\cite{venkatesh2008technology}, three about the four-panel comic (recall aid, ownership, autonomy), and two about the AI peer (friendliness, conversation quality). To support accessibility, each Likert point was paired with an emoji and a sentence label clarifying its meaning. We also asked which journaling method they preferred, offering several options with varying levels of AI collaboration.
    

    We initially planned optional interviews for adolescents; however, because parents preferred non-face-to-face participation, we instead collected written responses to six open-ended questions in the exit survey from adolescents who opted in. The questions mainly addressed reasons behind their ratings (\eg{}, intention to use, recall aid, friendliness, ownership, autonomy) and perceived differences between talking with a parent and the AI peer. Nine of the ten adolescents responded (all except \child{3}).

    For parents, the survey comprised 11 items: six TAM-based questions, two about the four-panel comic strip (content appropriateness, description coherence), and three about the AI peer (likability, facilitation of expression, conversation quality). Following the survey, parents participated in an hour-long interview with one researcher. The interview questions covered topics including their child's engagement with the app, their child's interactions with the AI peer and responses to the four-panel comic strip, the balance between parental involvement and their child's independence, observed changes in their child's expression, and parents' reflections on benefits, challenges, and desired improvements. All interviews were audio-recorded, anonymized, and transcribed for analysis.

\subsection{Data Analysis}
To characterize \sysname{} usage patterns, we conducted descriptive analyses of the collected logs for journal entries, and for duration and the number of turns per entry and by each phase (a \textit{turn} is a single exchanged message; \textit{adolescent turns} are participant utterances and \textit{system turns} are AI peer utterances). We also systematically coded the journal entries to analyze the distributions of key elements in terms of location, people, and activity.

\revised{We then quantitatively analyzed interaction logs between the autistic adolescent and the AI peer to examine how adolescents developed the panel contents over the course of a journaling session. For each component of the ABC-E framework (Antecedent, Behavior, Consequence, and Emotion), we identified (1) the turn in which the adolescent first mentioned the corresponding panel content and (2) the turn in which it was last revised, if the information changed more than once. We focused on when the first mention of and last revision of each panel content was made in each adolescent's journaling session, rather than the total number of turns taken. This was to analyze the more reliable markers of engagement, as interim revisions were often fragmented or distributed across multiple turns, preventing a reliable quantification of revisions.
We then examined patterns in the occurrence of these markers with respect to their relative positions across all turns in the session and the conversational phases in which they appeared (\phasearticulation, \phaseverification, \phaseelaboration, or \phaserevision).
In addition, we fit mixed-effects models~\cite{pinheiro2000mixed} to daily survey ratings of parental moderation and parent–adolescent positive conversation for assessing change over time.}

As for qualitative analysis, we analyzed parents' daily survey and the debriefing interviews using open coding and Thematic Analysis~\cite{braun2006using}. The first author generated initial codes and candidate themes, and then the entire research team discussed any disagreements and iteratively revised the themes to consensus. Through our comprehensive qualitative analysis, we revealed the multifaceted impact of \sysname{} on both adolescents and their parents, especially in terms of perceptions and conversations.

\subsection{Safety Protocols for Adolescents} 
\revised{
We implemented the following protocols in accordance with our approved IRB guidelines to ensure the safety of our adolescent participants.}
Each day during the two-week deployment, we reviewed the journal entries and the AI-adolescent exchanges to verify how the \revised{chatbot} prompted and guided responses, screening for potential risks (\eg{}, signals of distress, escalating conflict). If any such signals were observed, the first author would promptly engage the dyads and suspend the session. One of the co-authors, an autism expert and a certified youth counselor, served as a dedicated on-call counselor to provide immediate voice or video consultations. Where appropriate, a referral to a consulting psychiatrist for urgent care was also planned.

Adolescent participants were informed in advance that they could stop journaling at any time, and their parent was present in a safeguard role during system use. \revised{To prevent an excessive cognitive burden for our participants, we also instructed parents during the introductory session that if the chatbot seemed to burden their child by repeatedly prompting for elaboration on a particular aspect---a known erroneous behavior of LLMs~\cite{wei2024leveraging}---they could ask their child whether they wished to skip by saying ``\textit{move to next.}''}

\revised{No adverse event requiring counselor intervention occurred throughout the two-week deployment period.}



\section{Findings}
In this section, we present key findings from our deployment study \cameraready{in six parts.
In \autoref{sec:finding1}, we provide an overview of the overall system usage patterns observed during the two-week deployment.
In \autoref{sec:finding2}, we report on the interaction log analysis and discuss how adolescents disclosed and revised information within the ABC-E format.
Sections \ref{sec:finding3} through \ref{sec:finding5} primarily draw on exit surveys and debriefing comments.  
In \autoref{sec:finding3}, we describe how scaffolding and multimodal support guided adolescents in constructing narratives and facilitated their expression. 
In \autoref{sec:finding4}, we report on adolescents' engagement with and perceptions of the AI peer. 
In \autoref{sec:finding5}, we explore how parents perceived the impact of \sysname{} on their child's expression, independence, and parent--child interaction.
Lastly, in \autoref{sec:finding6}, we examine adolescents' and parents' perceived usefulness of \sysname{} and their intentions to adopt it in everyday life.}


\subsection{Overall System Usage}\label{sec:finding1}
Over the two-week deployment period, adolescents actively engaged in journaling with \sysname{}. Four out of 10 dyads used the system every single day, with no dyads skipping more than two consecutive days.
Adolescents recorded a total of 122 journal entries, averaging 12.2 entries per adolescent (\textit{SD} = 2.04; \textit{min} = 10 [\child{1}, \child{6}, \child{8--9}], \textit{max} = 15 [\child{5}]). Each session lasted 9~minutes and 43~seconds (\textit{SD} = 4m 33s) on average and involved 46.92 conversational turns with the system (\textit{SD} = 19.60), of which approximately 23 were adolescent turns. \revised{Mixed-effects models revealed no significant changes in session duration ($t = -1.20$, $p = .23$) or number of conversational turns ($t = -1.62$, $p = .11$) over the two-week period.}
During the sessions, adolescents spent most of the time in the \phaseelaboration{} phase (6m 17s with 25 turns) followed by the \phaserevision{} phase (1m 11s with 6.6 turns).

\revised{Adolescents journaled primarily at routine places (83 out of 122 entries; 68.03\%), most commonly at after-school classes ($N=40$) and at school ($N=25$).} 
%
From the activities described in the journal entries, we identified 13 activity types and grouped them into three high-level categories: \textit{leisure}, \textit{daily life}, and \textit{special activities} (see \revised{\autoref{appendix:categories} for an exhaustive list of categories}). On average, each adolescent contributed to 6.34 unique activity types (\textit{SD} = 1.69, \textit{min} = 4 [\child{8}], \textit{max} = 10 [\child{2}]). 
\revised{Almost half (46\%) of the journal entries incorporated \textbf{leisure} activities outside daily routines. Most adolescents frequently recorded \textit{sports and physical activities}, ranging from physical education classes (\eg{}, swimming, table tennis, taekwondo) to hobbies (\eg{}, bowling, badminton). Half of the adolescents recorded \textit{eating out}, typically emphasizing family bonding while dining together at restaurants. 
Adolescents also recorded a broad range of routine activities from their \textbf{daily lives} in 34\% of the entries ($N=42$), such as \textit{school activities}, \textit{domestic cooking and eating} with family or friends, and engaging in \textit{everyday activities} (\eg{}, walking or spending time at home with family). Some journal entries ($N=4$) also described \textit{social issues and conflicts} with friends and family.
One fifth of the journal entries ($N=24$) focused on \textbf{special activities}, rare and unusual events compared with everyday leisure, such as visiting an amusement park.}


\begin{figure*}[b]
  \centering
  \includegraphics[width=\textwidth]{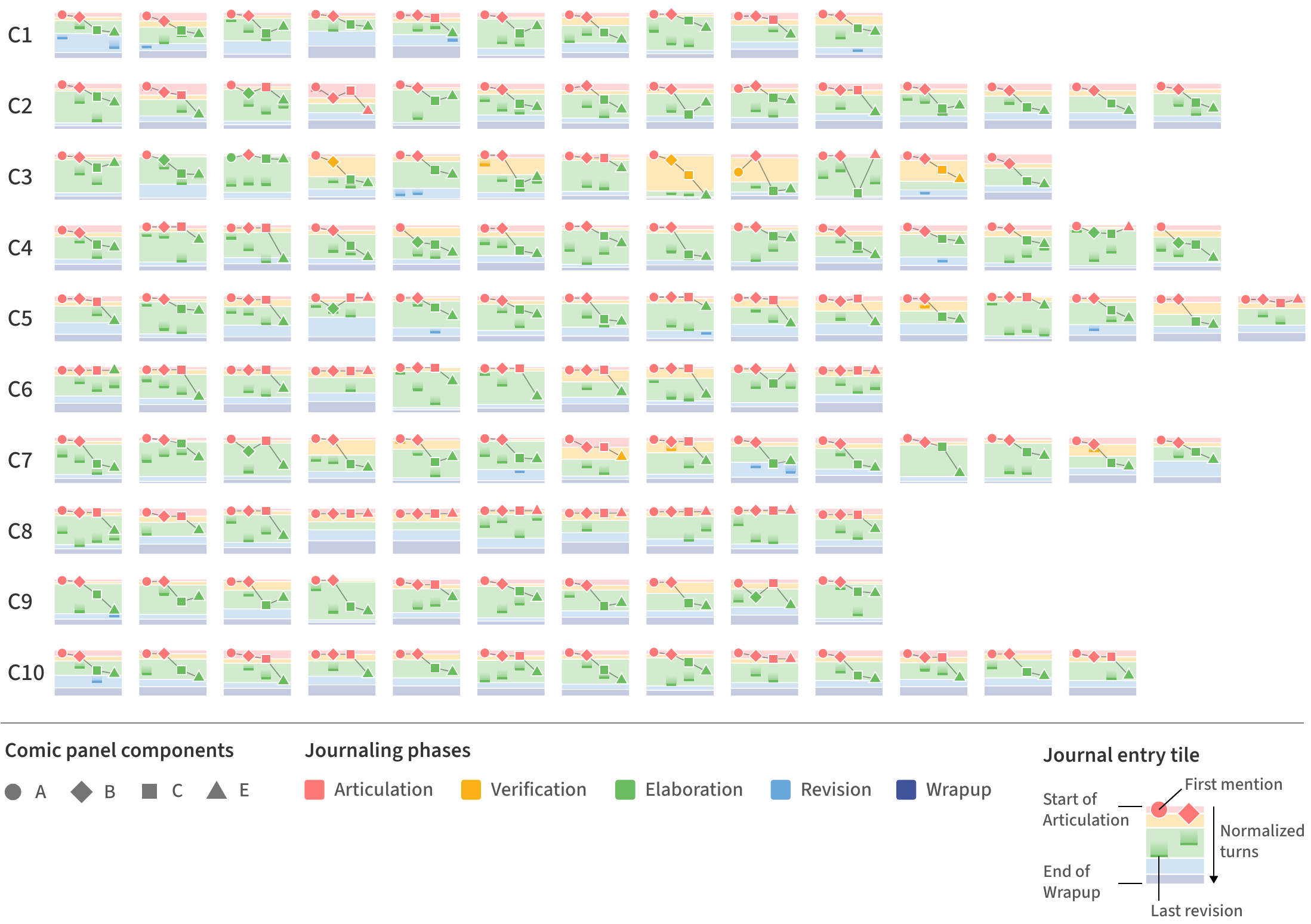}
  \caption{\revised{
  Normalized positions of first mentions and last revisions of the ABC-E information components across the conversational turns of each journal entry across all adolescent participants (\child{1--10}). Each visualization tile represents one journal entry. The color of each position marker indicates the journaling phases in which the mention occurred, while the background color of each journal entry tile denotes the actual normalized segments corresponding to each phase.}}
  \Description{A grid visualization showing normalized positions of first mentions and last revisions of ABC-E components across journal entries for 10 adolescent participants (C1--C10, rows). Each small tile represents one journal entry, with horizontal colored bands indicating journaling phases: Articulation (pink), Verification (yellow), Elaboration (green), Revision (light blue), and Wrapup (dark blue). Shape markers denote components---circle (A), diamond (B), square (C), triangle (E)---positioned where first mentioned and connected by lines to last revision points. A legend at bottom defines the shapes, phase colors, and tile structure (start of Articulation at top, end of Wrapup at bottom). Patterns vary across participants, showing individual differences in when and how ABC-E elements emerge during sessions.}
  \label{fig:individual_patterns}
\end{figure*}

\begin{figure*}[h]
  \centering
  \includegraphics[width=\textwidth]{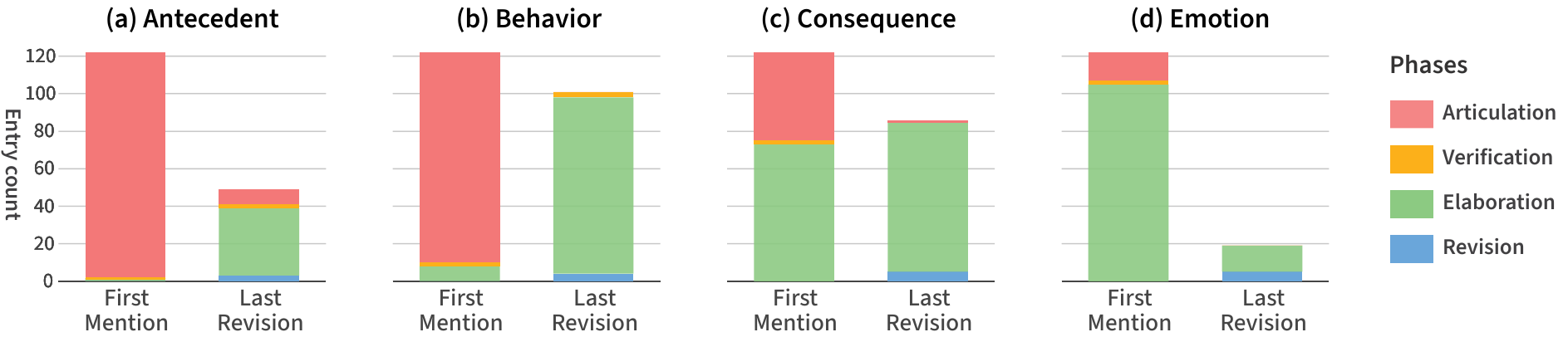}
  \caption{\revised{Distributions of the journaling phases in which each ABC-E component was first mentioned or last revised. The y-axis indicates the number of journal entries.}}
  \Description{Four stacked bar charts (a--d) showing distributions of journaling phases for first mentions versus last revisions of each ABC-E component. The y-axis indicates entry count (0--120). Colors represent phases: Articulation (pink), Verification (yellow), Elaboration (green), and Revision (blue). (a) Antecedent and (b) Behavior show first mentions predominantly in Articulation (tall pink bars ~120), while (c) Consequence and (d) Emotion have first mentions split between Articulation and Elaboration (mixed pink and green). Last revisions across all components shift toward Elaboration and Revision phases, with varying counts: Antecedent (~50), Behavior (~100), Consequence (~85), and Emotion (~20). The pattern indicates A and B emerge early, while C and E are often elicited during Elaboration.}
  \label{fig:mentions_by_abce}
  \labelphantom{fig:mentions_by_abce:antecedent}
  \labelphantom{fig:mentions_by_abce:behavior}
  \labelphantom{fig:mentions_by_abce:consequence}
  \labelphantom{fig:mentions_by_abce:emotion}
\end{figure*}

\subsection{\revised{Journaling Interactions}}\label{sec:finding2}

\begin{revisedenv}
Our interaction log analysis revealed a meaningful engagement pattern from the adolescent participants, mainly introducing the concrete details of an event, followed by reflections.
Of 488 comic panels in 122 journal entries, 232 cases were mentioned once (\ie{}, only the first mention exists) and the remaining 256 were at least once updated after being first mentioned. \autoref{fig:individual_patterns} visualizes the normalized positions of these 488 first mentions and 256 last revisions of the ABC-E information components across the conversational turns of each journal entry, and \autoref{fig:mentions_by_abce} summarizes the distributions of phases corresponding to first mentions and last revisions of each ABC-E component.
Adolescents appeared to exhibit a general tendency in when they disclosed and revisited their journal contents within the ABC-E framework, while the specific ways varied substantially across individuals. 

As shown in \autoref{fig:individual_patterns}, 
autistic adolescents often introduced \textit{Antecedent} and \textit{Behavior} early in their journaling processes.
Consistent with this observation, both components were typically first mentioned in the \phasearticulation{} phase---the first conversational phase---with the Antecedent appearing there in all but two journal entries (see \autoref{fig:mentions_by_abce:antecedent}) and Behavior in 92\% of them (see \autoref{fig:mentions_by_abce:behavior}). This pattern generally aligns with prior work, indicating that autistic individuals often foreground concrete, observable events~\cite{dindar2023autistic}.
In addition, because adolescents had already selected the setting and characters during the preceding \phasepreparation{} phase, they may have introduced the Antecedent-related information naturally in the subsequent phase.
Notably, although Behavior components were usually introduced early, they frequently underwent additional refinement as the journaling interaction progressed: most Behavior instances (83\%; $N=101$) were revised after their first mention, and 93\% of those revisions ended in the \phaseelaboration{} phase, during which the chatbot engages with adolescents to clarify incomplete information and elicit missing details for completing the comic panels (see \autoref{fig:mentions_by_abce:behavior}).
In contrast to these two components, \textit{Consequence} and \textit{Emotion} were typically introduced later, most often in the \phaseelaboration{} phase (60\% and 86\% of the first mentions of \textit{Consequence} and \textit{Emotion} instances, respectively; see \autoref{fig:mentions_by_abce:consequence} and \ref{fig:mentions_by_abce:emotion}). This suggests that the system's scaffolding may have supported adolescents in articulating components that are more reflective (\textit{Consequence}) or abstract (\textit{Emotion}) in nature, which they did not spontaneously provide in earlier stages.

    \end{revisedenv}

\subsection{Constructing Narratives with \sysname{}}\label{sec:finding3}
Based on the survey results and feedback from debriefing, we \revised{further} illustrate how \sysname{} guided adolescents' narrative construction through scaffolding and multimodal support.

    \ipstart{Stepwise Interaction Scaffolded Topic Selection and Narrative Construction}
    A majority of journaling sessions were guided by the scaffolds for selecting place and people. In 66.39\% of journal entries ($N=81$), adolescents opted in the selection mode, whereas they chose an open-ended option, \textit{I have something I want to write}, in 31.15\% of the entries ($N=38$). 
    Parents found the selection steps in the \phasepreparation{} phase as highly supportive as a starting point for their child. 
    Six parents
    emphasized that ``\textit{having clear options right in front of them really helped}'' (\parent{1}).
    Parents unanimously recognized the effectiveness of the step-by-step questioning in the ABC-E format for advancing the story. \parent{7} remarked, ``\textit{At first my child did not think concretely and just told things in his own way, but as he tried to answer the questions, he started to share more diverse, specific parts.}''
    Parents also noted that the composition of the ABC-E format helped their child grasp the essential structural pieces (\eg{}, who did what with whom, what happened, and then what feelings followed) of the situation (\parent{8}) and organize the narrative in temporal order (\parent{9}).
    Most parents (\parent{1}, \parent{4}, \parent{7--10}) emphasized that the \textbf{Emotion} part was particularly beneficial, encouraging their child to revisit how they felt and why. \parent{10} highlighted that this led to short but meaningful reflection: ``\textit{With a typical diary, my child might write, `there was an event and I felt good.' But it[\sysname{}] prompted her to think about what exactly triggered that feeling. [...]
    it prompted reflection on what the activities meant and how they made her feel.}''



\begin{figure*}[h]
    \centering
    \includegraphics[width=\textwidth]{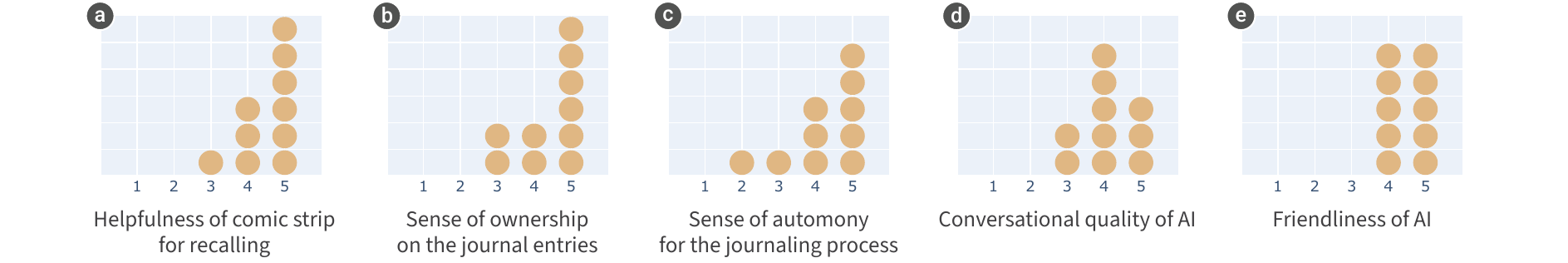}
    
    \caption{Distribution of adolescent participants' post-study ratings for \revised{helpfulness of comic strip for recalling (\circledigit{a}), the sense of ownership and autonomy (\circledigit{b} and \circledigit{c}), and the interactions with the AI peer (\circledigit{d} and \circledigit{e}). The x-axis denotes 5-point Likert scale ratings and} each circle represents the rating of an individual adolescent participant, stacked along the y-axis. For all scales, a higher number indicates a more positive rating.}
    \Description{The figure contains five small dot-plots (a--e), each showing adolescents' post-study 5-point ratings; the x-axis runs from 1 to 5 (higher = more positive). Each circle represents one participant's score; circles stack vertically to show multiple responses at the same value. Helpfulness of comic strip for recalling (a) is skewed toward 4--5, with a few mid-scale 3s. Ownership and autonomy (b--c): Sense of ownership of the journal entries and Sense of autonomy for the journaling process are concentrated at 4--5, with a few low scores. AI peer interaction (d--e): Conversational quality of AI and Friendliness of AI also cluster at 4--5, with occasional 3s and rarely any ratings at 1--2. All plots share the same grid and scale; distinctions are by panel labels and captions, not color.}
    \label{fig:survey_adolescent:feature}
\end{figure*}

    \ipstart{Visual Support and Speech Interaction Facilitated Reflection and Expression}
    In the exit survey, adolescents rated the comic strip as helpful for recalling the daily events ($M=4.5$, \textit{SD} = 0.71; see \autoref{fig:survey_adolescent:feature}-\circledigit{a}). They commented, for example, ``\textit{the drawing helped me think}'' (\child{1}) and ``\textit{seeing the drawing made it easier to understand}'' (\child{5}).
    Parents' observations supported this finding; eight of them reported that the comic strip served as an effective memory cue, helping children recall and elaborate on details. \parent{2} remarked, ``\textit{My child is not yet a fluent reader, so seeing the panels made it easier to recall and answer.}''

    Adolescents also expressed positive attitudes towards the voice modality, with seven (\child{1}, \child{3--7}, \child{10}) indicating a preference for `dialog-based comic journaling as currently implemented' other than `self-writing with AI-generated drawings' ($N=2$) or `fully independent' ($N=1$).
    Similarly, parents recognized that the voice-driven conversational modality was ``\textit{more accessible than writing}'' (\parent{1}), and two parents (\parent{2}, \parent{5}) noted that the method encourages verbalization and externalization of adolescents' thoughts. \parent{5} explained, ``\textit{When the transcribed words came out wrong, he swallowed and then tried to articulate more precisely, which seemed better for language development. With handwriting, a child can erase and tweak silently, but here the words have to be formed in the mind and produced aloud. Even if he hesitated, I could see him think again and try to fill the gaps, which I found very positive.}''


\subsection{Adolescents' Engagement with and Perceptions of the AI Peer}\label{sec:finding4}
We introduced a customizable, peer-like AI as a conversational partner to scaffold and guide journaling in \sysname{}.
Here, we report how the adolescents engaged with the AI peer in terms of customization and how they perceive it. 
As we incorporated significant AI guidance, we also investigated the sense of ownership and autonomy in the journaling process, as well as for the resulting journal entries.

\ipstart{Enthusiastic Customization of the AI Character}
Eight adolescents (80\%) heavily customized the AI peer's name, voice, and visual representation. The boys created a wide variety of characters: one \textit{male character} (\child{1}), one \textit{photorealistic male} (\child{7}), two \textit{photorealistic females} (\child{5--6}), and two \textit{stylized characters}, including a cartoon hero with a male voice and an octopus character with a female voice (\child{4}, \child{9}). All girls create characters with a female voice: one \textit{photorealistic female} (\child{3}) and one \textit{favorite bear character} (\child{10}). 
In debriefing, parents suspected that the customization feature appeared to have reinforced their child's engagement. \parent{5} described, ``\textit{We set the name to Thomas because my child liked it; 
[...] When the `lovely friend' appeared and said, `I'm Thomas,' my child just lit up and kept laughing---so the activity became more fun.}'' \parent{9} also remarked, ``\textit{
If there had been nothing---just `write'---it would have felt like homework, but having a favorite character made it feel distinctly different.}''

\ipstart{AI as a Friend}
According to the exit survey, adolescents generally perceived the AI peer as a `good friend' (\textit{M} = 4.5, \textit{SD} = 0.53; see \autoref{fig:survey_adolescent:feature}-\circledigit{e}). They commented that the AI as ``\textit{kind and friendly}'' (\child{1--2}, \child{5}, \child{10}), with a ``\textit{pretty voice}'' (\child{6}) and ``\textit{cute appearance}'' (\child{9}). They further noted that it ``\textit{spoke well to me}'' (\child{4}) and ``\textit{understood well when I explained}'' (\child{7}). Some adolescents also emphasized that the AI had ``\textit{enough human-like psychological understanding to comfort me}'' (\child{8}), and ``\textit{avoided prying into sensitive counseling details}'' (\child{10})---which contributed to making the interaction comfortable and non-intrusive.
Parents, in turn, shared the impression that their child seemed to treat the AI as a real friend. 
For example, \parent{3} noted, ``\textit{My daughter keeps asking whether [AI peer] is also in the fifth grade, and she says she wants to grow her bangs like [AI peer]'s.}'' 

\ipstart{Perceived Ownership and Autonomy in Collaborating with AI}
According to the exit survey, most adolescents retained a strong sense of ownership over the journal entries, with a high rating for the item `it felt entirely like my own journal' (\textit{M} = 4.4, \textit{SD} = 0.84; see \autoref{fig:survey_adolescent:feature}-\circledigit{b}). 
The few participants who gave a moderate score of 3 expressed a sense of shared ownership with their AI peer. \child{8} commented: ``\textit{In the finished diary, I could see [AI peer]'s effort and process to understand and transcribe my answers; it felt like collaboration.}''
Similarly, adolescents generally felt a high degree of autonomy, rating the item `I could write it in the way I wanted' favorably (\textit{M} = 4.2, \textit{SD} = 1.03; see \autoref{fig:survey_adolescent:feature}-\circledigit{c}). Exceptionally, \child{9} rated as not having a sense of autonomy, commenting ``\textit{Because the text and drawings did not turn out the way I wanted.}'' This highlights that for some users, a high degree of perceived control and predictability over the final output is crucial for autonomy.

\subsection{Parents' Perceived Impact of \sysname{}}\label{sec:finding5}
Based on the daily surveys for parents and the interviews, we explore how parents reflected on the tangible or potential changes that \sysname{} introduced for their child and themselves.

\ipstart{Expanding Adolescents' Vocabulary and Narrative Expression}
In debriefing, parents reported several key shifts in their child's patterns of conveying narratives over time when using \sysname{}. They noted improved central coherence (\parent{2}, \parent{4}, \parent{9}), as narratives moved from list-like sequences to converging on a single topic (\parent{9}). Additionally, adolescents began to internalize the ABC-E format (\parent{2--3}, \parent{6}, \parent{8}), shaping responses to fit it and producing more specific stories (\parent{8}). The topics also expanded from familiar family situations to new episodes from school (\parent{4}, \parent{10}), and similarly, parents reported a broader emotional range, with their child expressing more nuanced feelings (\parent{7}, \parent{10}), such as `confused' (\parent{7}) and `proud' (\parent{10}).

Some parents reported noticeable changes in everyday conversations and expressed surprise. \parent{7} pointed out ``\textit{In the last few days, my child explained situations more clearly. And when I misunderstood, he corrected himself---`not X, but Y'---in a way that felt like what the app had taught. In the past, he would have just said `it was Y', but now he points out the part he wants to fix. I realized that what we did here is starting to show up little by little in daily life, and that he is applying it. Even though the time was short, I felt it helped.}'' 
Other parents also described improvements such as more responsive answers to others' questions (\parent{4}) or longer conversations with more elaboration instead of one-word replies (\parent{5}). 

    \begin{figure}[t]
      \centering
        \centering
        \includegraphics[width=\linewidth]{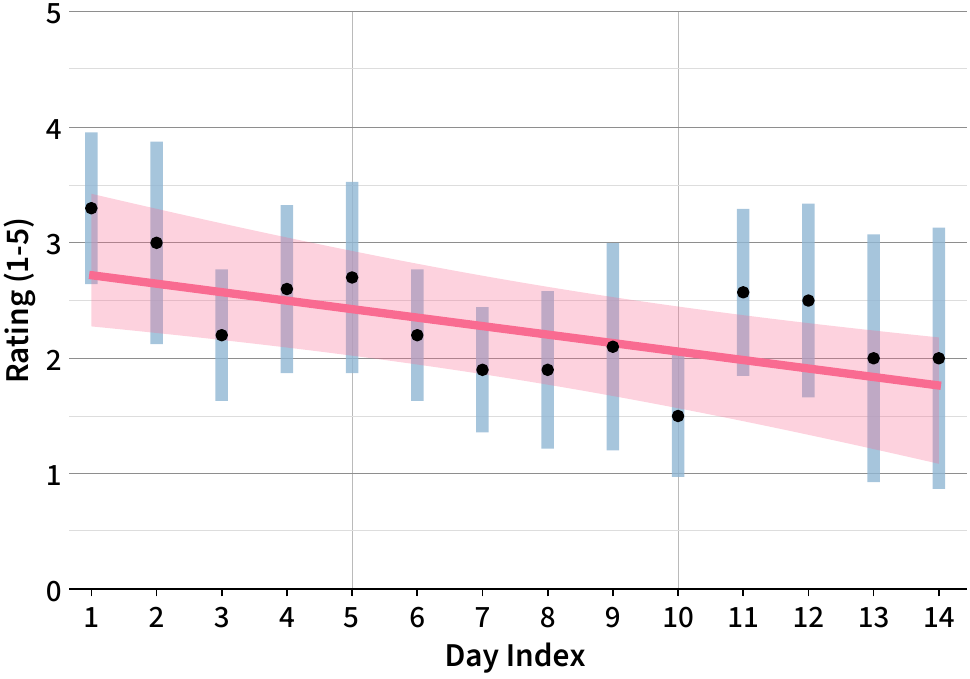}
        \caption{Daily trends in the estimated marginal means of parents' evaluation of the level of parental moderation during the journaling sessions over a 14-day period, after controlling for the random effect of individuals using mixed-effect models. The blue bars indicate the 95\% confidence interval.}
        \Description{A line-and-point chart shows daily estimated marginal means of parents' ratings of parental moderation during journaling over 14 days. The x-axis is Day (1--14) and the y-axis is Rating (1--5). For each day, a point marks the estimated marginal mean and a vertical error bar indicates its 95\% confidence interval. A fitted trend line with a surrounding confidence ribbon overlays the points. The fitted line decreases steadily from roughly above 3 on Day 1 to around 2 by Day 14, indicating a gradual downward trend across the study period. Confidence intervals vary in length by day and often overlap with adjacent days.}
      \label{fig:daily_trend:intervention}
    \end{figure}

    \ipstart{Shifting towards Independent Journaling}
    The mixed effects model analysis of the daily survey revealed that the level of parental moderation significantly decreased (\textit{p} = 0.001**; see \autoref{fig:daily_trend:intervention}) over time. In particular, eight parents responded that their child used \sysname{} independently without explicit support (rating = 1) on at least one day. This suggests that as the adolescents became familiar with \sysname{}, they demanded progressively less assistance from their parents to use it. 
    \parent{7} noted, ``\textit{Early on I intervened a lot, but midway through my child was taking over; toward the end it felt like the child was doing 90--100\%.}''
    In some cases, this independence evolved into self-directed engagement, where adolescents began to take the initiative themselves (\child{4}, \child{P9--10}). \parent{4} remarked: ``\textit{I initially set alarms on purpose to remind and urge my child to journal. But as he came to enjoy it, he began seeking it[\sysname{}] out and getting started on his own, which really surprised me.}''
    \begin{figure*}[b]
        \centering
        \includegraphics[width=0.9\textwidth]{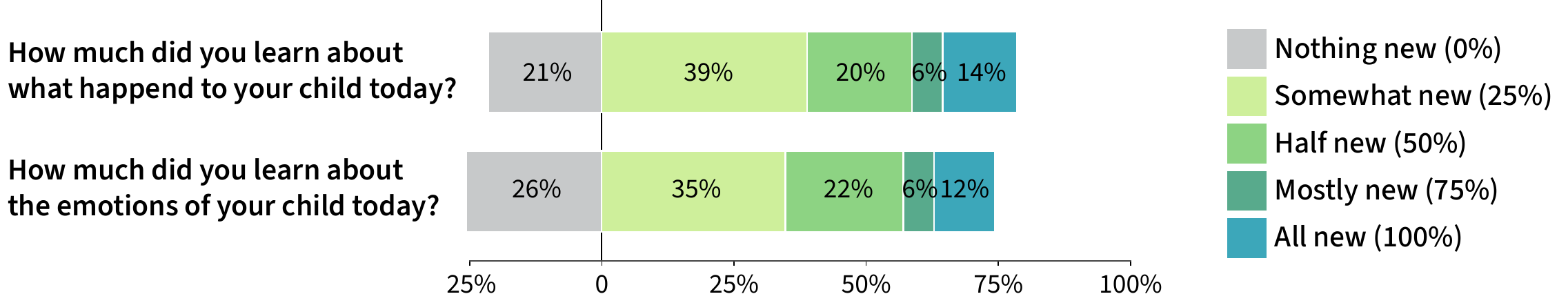}
        \caption{Distribution of parents' daily responses regarding the extent of new insights they gained about their child's daily events and emotions through \sysname{}.}
        \Description{Two horizontally stacked bars show parents' daily responses about how much new information they gained via AUTIVERSE. Each bar spans 0--100\% and is divided into five labeled segments: Nothing new (0\%), Somewhat new (25\%), Half new (50\%), Mostly new (75\%), and All new (100\%); percentages are printed on the segments.
        - What happened to the child today? The distribution is 21\% Nothing new, 39\% Somewhat new, 20\% Half new, 6\% Mostly new, and 14\% All new.
        - Emotions of the child today? The distribution is 26\% Nothing new, 35\% Somewhat new, 22\% Half new, 6\% Mostly new, and 12\% All new.
        For both questions, the modal response is "Somewhat new," followed by "Half new"; "All new" occurs less frequently than mid-level categories. Colors differentiate categories, but all categories are also text-labeled to avoid reliance on color alone.}
        \label{fig:event_emotion}
    \end{figure*}

    \ipstart{Enhancing Parental Awareness of their Child's Daily Experience}
    Our daily surveys captured the extent of new insights parents gained compared to what they already knew before reading the day's journal entries. \autoref{fig:event_emotion} illustrates the distributions of parents' daily responses by rating category. For both event-related insights (from the ABC panels) and emotion-related insights (from the E panels), parents reported gaining new insights on most days. As most of our parent participants accompanied their child to after-school activities and regularly received updates from school teachers, they were already familiar with many aspects of their child's daily events. However, parents noted that \sysname{} still uncovered previously unknown details about their child's specific daily events and emotions. For instance, \parent{6} discovered that her child's anxiety had led him to slam and break a keyboard, while \parent{5} learned that a seemingly fun trip to an amusement park had actually been a frightening experience.
    Parents also learned entirely new information that would not surface otherwise. \parent{8} mentioned ``\textit{It was nice to hear whether my child was doing well, eating properly, and what happened at school because those are times and spaces I cannot see.}''
    
    


    \ipstart{Fostering Parent-Adolescent Conversations}
    In daily surveys, parents reported that they had positive interactions with their child regarding \sysname{} ($M=3.99$, $SD=0.91$).
    Parents reported in the debriefing that sharing daily journal entries helped them (1) initiate everyday talk, (2) hold joint attention for deeper discussion, and (3) provide timely empathy and praise for their child.
    Five parents (\parent{1--3}, \parent{8}, \parent{10}) remarked that \sysname{} inspired everyday conversation topics for them to talk about their child's daily experience: ``\textit{Having this[\sysname{}] gave us a topic, we spent more time talking about it together. My child used to talk only about games, so it was nice that we could also talk about everyday life.}'' (\parent{2}).
    Further, \sysname{} provoked deeper conversations so that they can ``\textit{sit a bit longer and talk more deeply about the day}'' (\parent{5}).
    Two parents noted that the visual stories in \sysname{} helped them recognize opportunities to praise their child's effort and respond warmly. \parent{9} described, ``\textit{we use it as a tool to give that kind of praise like `You must have been tired from attending the class, but well done' or `Your writing has gotten much longer than before. Nice job.'}''    

\subsection{User Acceptance of \sysname{}}\label{sec:finding6}

    \begin{figure*}[t]
        \centering
        \includegraphics[width=\textwidth]{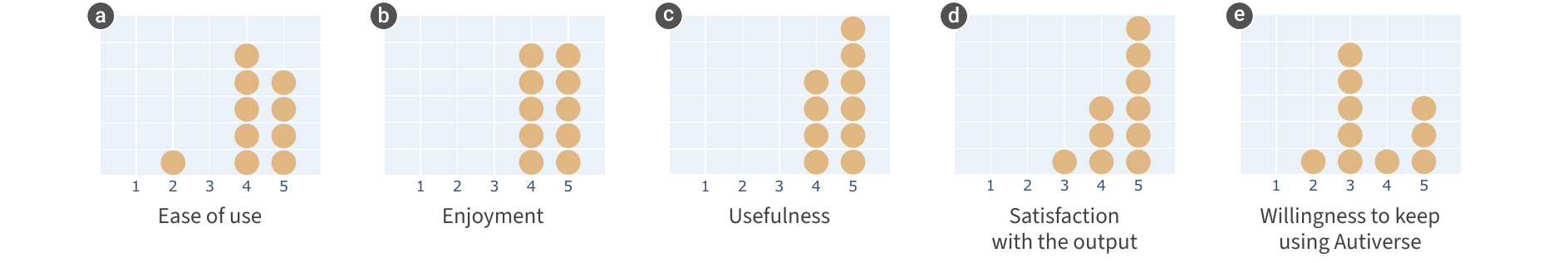}
        
        \caption{Distribution of adolescents' post-study ratings for utilities and benefits of \sysname{}. For all scales, the high number indicates a more positive rating.}
        \Description{The figure contains five small dot-plots (a--e), each showing adolescents' post-study 5-point ratings; the x-axis runs from 1 to 5 (higher = more positive). Each circle represents one participant's score; circles stack vertically to show multiple responses at the same value. Ease of use, Enjoyment, Usefulness, and Satisfaction with the output are skewed toward 4--5, with a few mid-scale 3s. Willingness to keep using AUTIVERSE shows a broader spread with several 3s alongside many 4s--5s. All plots share the same grid and scale; distinctions are by panel labels and captions, not color.}
        \label{fig:survey_adolescent:tam}
    \end{figure*}

    \begin{figure*}[t]
        \centering
        \includegraphics[width=\textwidth]{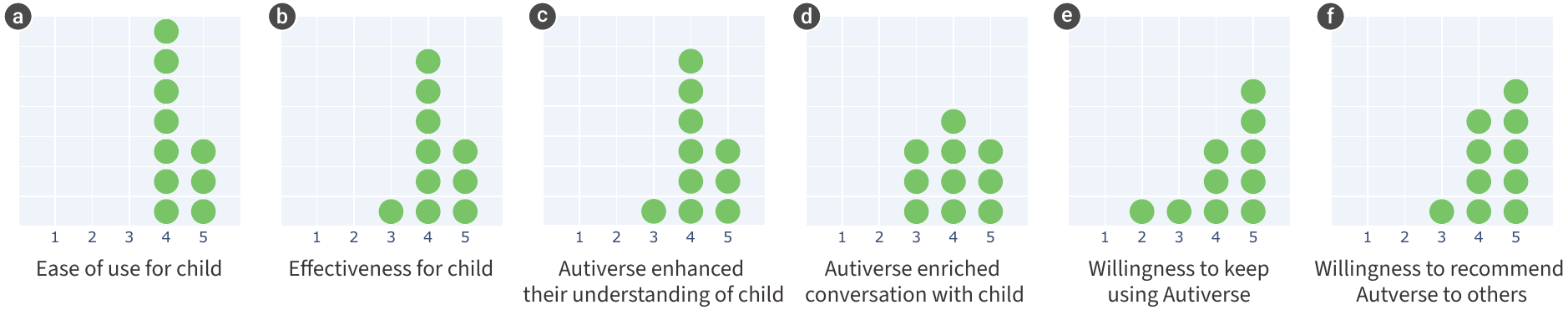}
        \caption{Distribution of parents' post-study ratings for utilities and benefits of \sysname{}. For all scales, the high number indicates a more positive rating.}
        \Description{Six small dot-plots (a--f) display parents' post-study 5-point ratings. In every panel, the x-axis is 1--5 (higher = more positive) and each circle is one parent, stacked vertically when multiple parents chose the same value.
    (a) Ease of use for child: ratings cluster at 4--5, indicating parents generally felt the system was easy for their child to use.
    (b) Effectiveness for child: mostly 4--5 with an occasional lower score (≈2--3).
    (c) AUTIVERSE enhanced their understanding of child: concentrations at 4--5, with a few 3s.
    (d) AUTIVERSE enriched conversation with child: values are centered around 3--4, with fewer at 5, suggesting comparatively more mixed views here.
    (e) Willingness to keep using AUTIVERSE: primarily 4--5 with a small number of 2--3 responses.
    (f) Willingness to recommend AUTIVERSE to others: again 4--5 dominate, with occasional 2--3.
    All panels share identical scales and rely on panel labels (not color) to distinguish constructs; the overall pattern shows generally positive evaluations, with conversational enrichment (d) being somewhat lower than other items.}
        \label{fig:survey_parent}
    \end{figure*}

\revised{In the exit survey, adolescent participants rated \sysname{} positively across ease of use, enjoyment, usefulness for recounting events, and the likability of the completed comic strip} (see \autoref{fig:survey_adolescent:tam}, \circledigit{a}--\circledigit{d}). In contrast, they gave mixed ratings on their willingness to keep using \sysname{} (\autoref{fig:survey_adolescent:tam}-\circledigit{e}). \revised{Those who rated 3 (`not sure') or below described resistance to diary-keeping} (\eg{}, ``\textit{It's fun, but I don't like writing a diary}''-\child{1}) or \revised{anticipatory anxiety about sustaining the activity} (\eg, ``\textit{If I keep doing it, there might be times when it feels hard.}''-\child{2}), a pattern aligned with autism-related anxiety~\cite{van2011anxiety}.
Parent participants also showed high acceptance of \sysname{} (see \autoref{fig:survey_parent}). \revised{They perceived \sysname{} as easy for their child to use and effective in supporting the child's expression and helping them understand the child's thinking. Parents also reported that \sysname{} enriched parent–adolescent conversations and expressed high willingness to adopt \sysname{} as well as to recommend it to other parents.} Motivated by their experience in the study, three parents expressed their intention to continue the journaling activity, with \parent{2} already having begun the practice: ``\textit{This became the trigger for us to write a journal every day! So last Saturday, we bought a bunch of stickers, since my child said she wanted to write while sticking on various characters and objects.}'' 

\section{Discussions}
\cameraready{In this section, we reflect on our findings to provide broader implications for designing AI-guided journaling systems that support the unique needs of autistic adolescents. In \autoref{sec:discuss7.1}, we discuss how \sysname{}'s scaffolding alleviated executive challenges in narrative construction. In \autoref{sec:discuss7.2}, we reflect on how developmental readiness and autism-specific traits might shape adolescent participants' engagement and their varying needs for parental support. In \autoref{sec:discuss7.3}, we suggest a hybrid approach for adaptive scaffolding that combines data-driven recommendations with parental customization to accommodate the wide variability in autism profiles. In \autoref{sec:discuss7.4}, we address potential socio-emotional risks in peer-like AI design and suggest strategies to mitigate them. In \autoref{sec:discuss7.5}, we envision a balanced approach that transitions from mediated to independent system use that fosters autistic adolescents' self-directed reflection while maintaining safety. Lastly, in \autoref{sec:discuss7.6}, we discuss limitations of our study that may affect the generalizability and interpretation of findings as well as future work.}

\subsection{Scaffolding Narrative Construction with Multimodal AI for Autistic Adolescents}\label{sec:discuss7.1} 
\sysname{} supported autistic adolescents in initiating and organizing their daily narratives by scaffolding the journaling process. Initial scaffolds, such as selecting a location and people, addressed the common challenge of deciding what to talk about due to executive dysfunction in autism~\cite{hill2004executive}. Step-by-step, question-based phases also mitigated the cognitive load of journaling by breaking the narrative process into manageable questions~\cite{petersen2014systematic}, enabling adolescents to populate the ABC-E format successfully. 
\begin{revisedenv}
However, the ABC-E components differed in how much scaffolding they required. While \textit{Antecedent} and \textit{Behavior} emerged readily during early conversation---consistent with autistic individuals' tendency to focus on concrete, observable details~\cite{dindar2023autistic}---\textit{Consequence} and \textit{Emotion} required more substantial scaffolding, with most first mentions occurring during the \phaseelaboration{} phase when the chatbot actively probed for these elements. This pattern suggests that components involving causal reasoning and emotional introspection are less immediately accessible and can benefit from explicit prompting. The high revision rate for \textit{Behavior} components---predominantly during the \phaseelaboration{} phase---further illustrates the iterative nature of narrative construction, with adolescents meaningfully engaging with clarifying questions to add specificity rather than treating them as intrusions. 
These findings carry important implications for designing conversational AI systems for autistic populations. Rather than attempting to elicit a full narrative at once, systems should anticipate that some components will emerge naturally while others require a more deliberate scaffolding. The progressive internalization of the ABC-E format in our study suggests that a structured format, when combined with adaptive conversational support, can function as a scaffold that effectively guides adolescents to identify and learn how to unpack key narrative components (\eg{}, consequence, emotion) in their daily lives, gradually helping them better express themselves in a more coherent narrative over time.
\end{revisedenv}

\subsection{\revised{Developmental and Autism-related Characteristics Shaping Engagement and Autonomy}}\label{sec:discuss7.2} 
\begin{revisedenv}
While eight autistic adolescents achieved independent use of \sysname{}, the remaining two required parental assistance throughout the study. In addition, although adolescents generally found the activity enjoyable, some also expressed resistance to continued use. 
Although our sample size does not provide evidence that age is a consistent predictor of autonomy---adolescents aged 13--14 showed variation in their reliance on parental support---contrasting cases in our sample suggest that developmental readiness may play a role. One of the two adolescents who required consistent parental assistance throughout the study was also the youngest participant (\child{3}; 11-year-old elementary school student), whereas those who progressed to self-initiated journaling include the oldest participant (\child{10}; 17-year-old high school student). These patterns align with prior work indicating that metacognitive awareness and autonomous motivation tend to increase with age~\cite{steinberg1986vicissitudes}.
Second, co-occurring conditions such as anxiety disorders---prevalent in 40–84\% of autistic youth~\cite{sukhodolsky2013cognitive}---may contribute to anticipatory resistance even to positive experiences, as reflected in \child{2}'s earlier remarks: ``[...] \textit{there might be times when it feels hard}.''
Third, autism-specific traits---such as perfectionism~\cite{greenaway2010dysfunctional} or need for control~\cite{chaxiong2022restricted}---may also have shaped how adolescents engaged with or interpreted the system's output. For instance, \child{9} rated his sense of autonomy low (2/5), noting that the output did not quite match his expectations. 
Therefore, the scaffolded narrative construction in \sysname{} may be particularly effective for older adolescents with lower anxiety and greater cognitive flexibility, whereas identifying effective ways to support younger or more perfectionistic adolescents warrants future research.
\end{revisedenv}

\subsection{\revised{Designing Adaptive and Customizable Scaffolding for Richer and Sustained Journaling}}\label{sec:discuss7.3} 
\begin{revisedenv}
During the debriefing, one parent reported that their child gradually disengaged from journaling as the child's skills advanced beyond the system's fixed step-by-step structure.
To accommodate the wide variability in autism profiles and developmental readiness, future systems may benefit from a hybrid approach that combines parental cues with data-driven recommendations derived from adolescent performance. By analyzing conversational data and journal content (\eg{}, vocabulary use, sentence complexity, or the range of emotions expressed), the system could recommend an appropriate level of scaffolding or interaction strategies that account for the adolescent's current status. In addition, prior research on interactive systems in autism contexts has consistently emphasized the importance of supporting rich \textit{customization} and \textit{personalization} of system features~\cite{Aguiar2022autismguide,park2025autismreal,Sharmin2018AutismTech}. Building on the system's recommendations, parents can further tailor the journaling experience for the autistic adolescents. 
For example, an adolescent who struggles to complete four panels might begin with a simpler narrative structure, prioritizing concrete facts (\eg, BC or ABC) before progressing to the full framework. More advanced adolescents could try extended narrative structures with additional panels; however, exploring the effective extension of the narrative framework in autism contexts remains an open research question. \needtocheck{}


\end{revisedenv}

\subsection{\revised{Addressing Socio--Emotional Risks in Peer-like AI Design}}\label{sec:discuss7.4} 
\begin{revisedenv}
Our study showed the promise of positioning an AI peer as a collaborative, non-didactic partner that is customizable to user preferences, enabling adolescents to interact with it in an enjoyable, friend-like manner.
Despite these positive reactions, the peer-like AI design may also pose risks if used over extended periods, particularly in shaping autistic adolescents' social interaction.

\textit{Emotional dependency} is an emerging issue in AI companionship research~\cite{visioemotional,papadopoulos2025use}. For autistic adolescents who often struggle to build peer relationships~\cite{bauminger2000loneliness}, AI companions may become a primary source of their social-emotional validation~\cite{papadopoulos2025use}. Although our two-week study did not reveal problematic attachment, indicators such as anthropomorphic questions (\eg{}, \child{3}'s inquiry about the AI's school grade) warrant caution~\cite{seo2024chacha}.
Also, LLMs tend to exhibit \textit{sycophancy}, excessively validating users~\cite{malmqvist2025sycophancy,cheng2025social}. \sysname{}'s AI peer was designed to be supportive and receptive, but overly accepting the user across prolonged interactions might inadvertently reinforce maladaptive interpretations.
Furthermore, AI's \textit{idealized conversational behaviors} (\eg{}, infinite patience, consistent interest, and empathy) differ from real human interactions, where others may be distracted, dismissive, or uninterested. This mismatch may foster unrealistic expectations in autistic adolescents and hinder their ability to navigate social friction.

To mitigate these risks, peer-like AIs need mechanisms for gentle disagreement when adolescents express ethically or morally inappropriate content (\eg{}, hostile attributions toward others, self-deprecating statements). In addition, when adolescents articulate experiences using aggressive or socially inappropriate language, the AI could suggest alternative, constructive phrasings through linguistic scaffolding, while preserving emotional authenticity. For instance, if an adolescent inputs ``\textit{I hate my stupid teacher who always ignores me,}'' the system could offer a reframing such as ``\texttt{It sounds like you felt frustrated when your teacher didn't notice you.}'' In this way, the system could support both emotional regulation and socially appropriate communication. 
\end{revisedenv}



\subsection{\revised{Balancing Adolescent Autonomy and Parental Mediation}}\label{sec:discuss7.5} 
\begin{revisedenv}
While we designed \sysname{} for independent use, we involved parents in the deployment study for safeguarding and potential troubleshooting. Although the adolescent participants shared a wide range of personal narratives---often revealing new information to parents---their presence may have influenced the depth of sensitive disclosures.
For example, adolescents may have been less inclined to discuss peer conflicts, romantic interests, or negative emotions toward family members in \sysname{}. As adolescents grow toward greater autonomy, supporting a shift to more independent use becomes developmentally appropriate.
In our study, some adolescents began to engage with \sysname{} on their own initiative, aligning with developmental theories of autonomy and individuation~\cite{blos1967second,larson1997emergence,steinberg1986vicissitudes}. This progression suggests that the scaffolded structure of \sysname{} may facilitate a gradual move from mediated to independent practice. 
Future systems could incorporate adaptive mechanisms that dynamically adjust parental involvement according to adolescents' demonstrated competence and safety indicators. After an onboarding period of supervised use, the system may introduce `independence milestones'---sessions in which adolescents journal alone while parents remain available but not present. These sessions could be gradually extended in frequency and duration, supported by stepwise rewards for adolescents and AI-generated summaries for parents that emphasize safety-related signals rather than full journal content~\cite{seo2024chacha,seo2025arch}. Such a phased approach may preserve necessary safeguarding while cultivating adolescents' capacity for private reflection and expressive independence. 
Future research should examine how different levels of parental involvement shape adolescents' disclosure patterns and identify reliable indicators of readiness for fully independent use across diverse cognitive and developmental profiles within the autism spectrum.
\end{revisedenv}

\subsection{Limitations and Future Work}\label{sec:discuss7.6} 
In this section, we discuss the limitations of our study that could impact the generalization of the findings.
First, our study utilized an LLM predominantly trained on Western languages to support autistic adolescents in a Korean context. While the AI's outputs were comprehensible to participants, we observed that LLMs took longer to generate responses due to inefficiencies in tokenizing Korean compared to English~\cite{ahia2023all}. This highlights an opportunity for future work to employ an LLM specifically fine-tuned on the Korean language and culture to enhance system responsiveness, potentially leading to greater user engagement.

Second, although we aimed for diversity (\eg{}, adolescents' gender, age, communication characteristics), all participants were residents of South Korea, a country with high AI adoption~\cite{ART003179482}. Furthermore, while our study with 10 autistic adolescents revealed promising opportunities for \sysname{}, this cohort may not be representative of the entire autism spectrum. Therefore, future research with a more diverse population---varying in socio-cultural context, AI literacy, and representation across the autism spectrum---would be valuable for exploring the system's broader applicability and uncovering different use patterns.



Third, our reward-based feature was limited to a simple mechanism of awarding three stamps, lacking more sophisticated mechanics. During the debriefing interview, parents reported that their child particularly enjoyed the popping animation of the stamps. However, they emphasized that more game-like elements, such as scores or rankings, would be crucial for substantially boosting their child's long-term engagement. Therefore, future work should focus on designing and integrating these advanced gamification features to create a more compelling and sustainable user experience.

\revised{Lastly, parental involvement or system novelty may have influenced the overall user experience. Since this work is one of the first steps in designing and developing an AI-guided journaling system accessible to autistic adolescents, our primary aim was to explore the feasibility of our approach while ensuring participant safety. Future research can incorporate comparative experimental designs to more rigorously disentangle system effects from contextual influences. Given that autism research requires particular attention to equity and participant burden, and that ethical concerns might arise from a conventional control group without access to the system, especially when parents may perceive it as a supportive intervention, future work may benefit from a waitlist-control design~\cite{Valentine2024WaitingList}, which offers a more ethically appropriate comparison while enabling rigorous evaluation.}

\section{Conclusion}
We presented \sysname{}, an AI-guided multimodal journaling app that scaffolds narrative construction of autistic adolescents. Informed by formative interviews with professionals and parents of autistic adolescents, \sysname{} incorporates a peer-like AI that elicits key details in the ABC-E format via stepwise dialogue and transforms them into an editable four-panel comic strip. Through a two-week deployment with 10 adolescent--parent dyads, \sysname{} proved to be a feasible and effective tool for helping adolescents organize their experiences and emotions into more coherent narratives and created a comfortable, enjoyable space for sharing. Parents reported learning additional details that often go unnoticed in routine conversations, enabling more meaningful follow-ups at home. Our findings highlight the promise of combining conversational prompts with visual supports to lower the executive burden of journaling. 
\revised{Specific areas of future work include providing adaptive scaffolding to accommodate diverse autism profiles and desiginng socio-emotionally appropriate AI peers.
In closing, we believe that \sysname{} can empower autistic adolescents as well as their caregivers, bridging the communicative gaps in sharing their daily events and emotions.}

\begin{acks}
We thank our participants from the formative interviews and the deployment study for their time and effort. We are also grateful to Jungeun Lee for providing feedback on our paper draft. This work was supported through a research internship at NAVER AI LAb of NAVER Cloud.
\end{acks}

\bibliographystyle{ACM-Reference-Format}
\bibliography{bibliography}

\newpage
\onecolumn
\appendix

\section{Generative Pipelines of \sysname{}}\label{appendix:pipeline}
\begin{revisedenv}
In the journaling process with \sysname{}, all phases except \phasepreparation{}, which consists of only touch inputs, incorporate generative pipelines with LLM-infused components for conversation, dialogue analyses, and structured information generation. \autoref{fig:pipeline:description} illustrates the generative pipeline flows of phases from \phasearticulation{} to \phasewrapup{}.

\begin{figure*}[b]
    \centering
    \includegraphics[width=\textwidth]{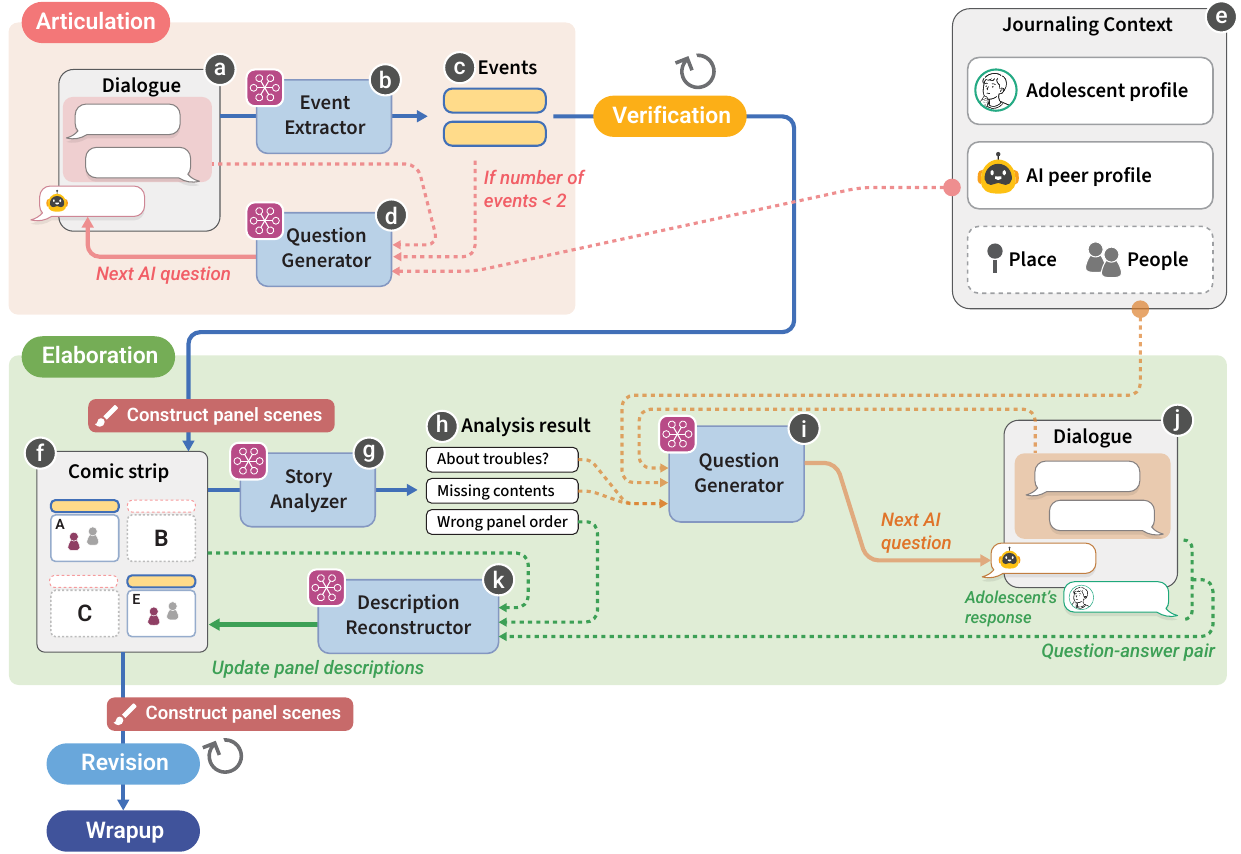}
    \caption{Generative pipeline flows of \sysname{}. In the \phasearticulation{} phase, the \circledigit{b} Event Extractor identifies \circledigit{c} events from the \circledigit{a} dialogue. If fewer than two events are found, the \circledigit{d} Question Generator formulates a follow-up question. The process then moves to the \phaseverification{} phase, which applies user modifications to the preliminary descriptions. After \raisebox{-2pt}{\includegraphics[width=9pt]{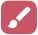}} constructing panel scenes based on these descriptions, the \phaseelaboration{} phase begins, iteratively updating the \circledigit{f} panel descriptions. In this loop, the \circledigit{g} Story Analyzer checks the descriptions for issues, and \circledigit{h} its results guide the \circledigit{i} Question Generator to ask for more details. The user's response, captured in the \circledigit{j} dialogue, is used by the \circledigit{k} Description Reconstructor to update the panel descriptions. Finally, after \raisebox{-2pt}{\includegraphics[width=9pt]{figures/icon_scene.pdf}} the panel scenes are constructed once more, the process concludes with the \phaserevision{} and \phasewrapup{} phases.}
    \Description{The figure shows AUTIVERSE's generative pipeline as a left-to-right workflow with phases and iterative loops. In the Articulation phase (orange, top), the adolescent's Dialogue (a) feeds an Event Extractor (b) that outputs candidate Events (c); if fewer than two events are detected, a Question Generator (d) issues a follow-up prompt. A side panel labeled Journaling Context (e) supplies profile, place, and people information that conditions question generation. After Verification, the process enters Elaboration (green, bottom): the system constructs panel scenes (f) for a provisional comic, then a Story Analyzer (g) inspects descriptions and produces Analysis results (h) (e.g., "about troubles?", "missing contents," "wrong panel order"). These results guide another Question Generator (i) that asks for more details in Dialogue (j); the adolescent's response is used by a Description Reconstructor (k) to update panel descriptions. Dashed arrows indicate an iterative loop across g–k, after which the system re-constructs scenes (f). The pipeline concludes with Revision and Wrap-up stages.}
    \label{fig:pipeline:description}
\end{figure*}

\subsection{Identifying Main Event}\label{gp:main_event}
    In the \phasearticulation{} phase, the chatbot aims to elicit at least two pieces of events from the adolescent. For each turn, the current dialogue (\circledigit{a} in \autoref{fig:pipeline:description}) is analyzed by \textbf{Event Extractor} (\circledigit{b} in \autoref{fig:pipeline:description}), which identifies pieces of events (\circledigit{c} in \autoref{fig:pipeline:description}; \eg{}, ``\textit{I played with Oliver.}'' and ``\textit{Oliver felt bad.}'' extracted from adolescent message, ``\textit{I played with Oliver and he felt bad.}'').
    The \textbf{Question Generator} (\circledigit{d} in \autoref{fig:pipeline:description}) generates the next AI message considering the current dialogue and other contexts (\circledigit{e} in \autoref{fig:pipeline:description}), including the adolescent's profile (\eg{}, age, gender, interests), customized AI peer profile, and the place and people set in the \phasepreparation{} phase.
    These contexts ensure that the generated questions are considerate of the adolescent's cognitive profile, allowing the AI to acknowledge off-topic responses appropriately, while gently guiding the conversation back to the day's main events.

    \subsection{Constructing Panel Descriptions}
    In the \phaseelaboration{} phase, the system iteratively updates the panel descriptions of the comic strip (\circledigit{f} in \autoref{fig:pipeline:description}) to organize a journal entry in the ABC-E format. At each iteration, the \textbf{Story Analyzer} (\circledigit{g} in \autoref{fig:pipeline:description}) inspects the current panel descriptions based on three criteria (\circledigit{h} in \autoref{fig:pipeline:description}): (1) Whether the main events involve troubles such as social or emotional conflict; (2) whether there are missing or underdeveloped elements (\eg{}, missing actors, unclear actions, or insufficient emotional description) for each panel description; and (3) whether each panel description adheres to the ABC-E format and whether the descriptions within a panel present events in chronological order. If the \textbf{Story Analyzer} determines that all panel descriptions are complete and correctly ordered, the system proceeds to the next phase (\phaserevision{}).

    Otherwise, the \textbf{Question Generator} (\circledigit{i} in \autoref{fig:pipeline:description}) generates the next AI question to elicit the missing elements from the adolescent (\eg{}, ``\texttt{What happened to Oliver that made him feel bad?}''). Similar to \textbf{Question Generator} (\circledigit{d} in \autoref{fig:pipeline:description}) in the \phasearticulation{} phase, it considers the current dialogue (\circledigit{j} in \autoref{fig:pipeline:description}) and other contexts (\circledigit{e} in \autoref{fig:pipeline:description}), along with the analysis results on troubles and missing content (\circledigit{h} in \autoref{fig:pipeline:description}). If the current events involve troubles, the questions focus on \textit{why} it occurred; otherwise, they focus on \textit{how} it unfolded. 
    The AI questions are typically open-ended or phrased to provide options verbally (\eg{}, ``\texttt{How did you use the Oliver eraser? Did you just erase it quickly, or did you do something else too?}''), but when asking about emotions, considering autistic adolescents' common challenge to express emotions explicitly~\cite{lartseva2015emotional}, the system presents a list of buttons for 12 emotions to choose from: \textit{joyful, glad, happy, excited, sad, angry, upset, scared, afraid, surprised, amazed,} and \textit{bored}. This set is a subset of Plutchik's Wheel of Emotion~\cite{plutchik1980general}, curated in consultation with one author---an autism expert---to exclude emotions that are rarely understood by autistic youth.

    After receiving the adolescent's response, the \textbf{Description Reconstructor} (\circledigit{k} in \autoref{fig:pipeline:description}) updates the panel description using the latest question-answer pair (\circledigit{j} in \autoref{fig:pipeline:description}) and the analysis results on panel order (\circledigit{h} in \autoref{fig:pipeline:description}).
    It first rearranges panel elements to ensure chronological consistency, then integrates new content into the appropriate panel based on narrative flow. Finally, it performs sentence cleanup and tense normalization to ensure that all content is in grammatically correct first-person past tense.

    \subsection{Applying Modifications and Wrapping Up}
    In the \phaseverification{} phase and the \phaserevision{} phase, adolescents can revise the generated panel descriptions by verbally requesting modifications (\eg{}, correcting or adding details). Then, the system incorporates these changes into the corresponding panel descriptions.
    
    When the adolescent confirms that no further revisions are necessary in the \phaserevision{} phase, the process advances to the \phasewrapup{} phase, which finalizes the journaling. This phase executes two sequential generative pipelines based on the final panel description: (1) the system generates a warm and personalized response to the adolescent, providing encouragement and emotional closure; and subsequently (2) proposes three candidate titles designed to be emotionally resonant and contextually appropriate, thereby concluding the journaling process with a title.
    
\subsection{\raisebox{-2pt}{\includegraphics[width=12pt]{figures/icon_scene.pdf}}~Constructing Panel Scenes}

Visual scenes of the panels are initially generated at the beginning of the \phaseelaboration{} phase and are kept updated during the \phaserevision{} phase (see \raisebox{-2pt}{\includegraphics[width=12pt]{figures/icon_scene.pdf}} in \autoref{fig:pipeline:description}).
\autoref{fig:pipeline:scene} illustrates the generative process of constructing scene information from the panel descriptions. The scenes are represented in a parametric JSON format, so the pipeline incorporates only text generation. Receiving panel descriptions (\circledigit{a} in \autoref{fig:pipeline:scene}), the \textbf{Element Extractor} (\blackrectsmall{1} in \autoref{fig:pipeline:scene}) identifies essential elements in each panel (\circledigit{b} in \autoref{fig:pipeline:scene}): actors along with associated attributes (actions, dialogue lines, thoughts, and emotions); objects or concepts (\eg, `ball', `eraser', `cooking'); and the settings (\eg, `Classroom'). From this information, the \textbf{Topology Calculator} (\blackrectsmall{2} in \autoref{fig:pipeline:scene}) formulates the adjacency relationship among actors, objects, and concepts, for example, imposing the constraint that `Oliver' and `Teacher' should be placed side by side (\circledigit{c} in \autoref{fig:pipeline:scene}).
Finally, the \textbf{Element Placer} (\blackrectsmall{3} in \autoref{fig:pipeline:scene}) determines the actual coordinates of each element on a $5\times5$ grid in accordance with these relationships. Actors' attributes such as dialogue lines and emotions are rendered on the client side in close proximity to the corresponding actors (\circledigit{d} in \autoref{fig:pipeline:scene}).

\begin{figure*}[t]
    \centering
    \includegraphics[width=\textwidth]{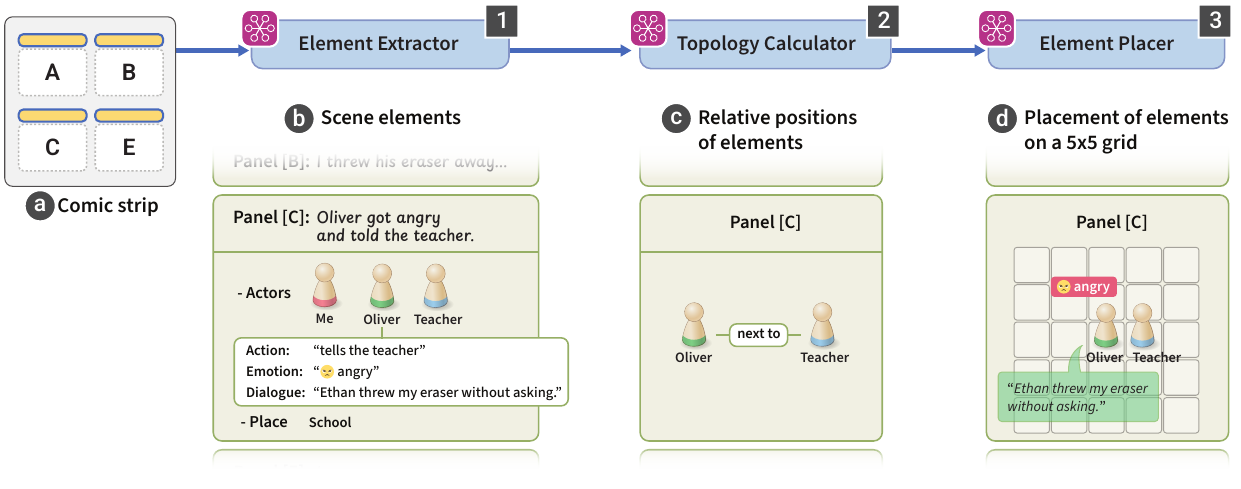}
    \caption{Generative pipeline for constructing panel scenes. The \blackrectsmall{1} Element Extractor first identifies \circledigit{b} scene elements from \circledigit{a} each panel description. Then the \blackrectsmall{2} Topology Calculator determines \circledigit{c} their relative positions of elements. Finally, the \blackrectsmall{3} Element Placer generates the \circledigit{d} placement of these elements on a 5$\times$5 grid to complete the scene.}
    \Description{The figure illustrates the panel-scene construction pipeline in three sequential modules with arrows from left to right: (1) Element Extractor, (2) Topology Calculator, and (3) Element Placer. A small 2×2 comic strip at left marks panels A, B, C, and E; the example proceeds with panel C. Under (1), the Scene elements (b) card lists items parsed from the panel-C description: Actors (icons for Me, Oliver, Teacher), Action "tells the teacher," Emotion "angry," the Dialogue text "Ethan threw my eraser without asking," and Place "School." Under (2), the Relative positions (c) card specifies spatial relations between elements, e.g., Oliver next to Teacher. Under (3), the Placement (d) card shows the elements laid out on a 5×5 grid: Oliver and Teacher appear side-by-side; the "angry" tag attaches to Oliver; the dialogue bubble is anchored in the panel. The pipeline outputs the completed panel-C scene while preserving labels so information is not conveyed by color alone.}
    \label{fig:pipeline:scene}
\end{figure*}

\end{revisedenv}

\newpage
\section{Additional Results from Deployment Study}
\revised{In this section, we share additional details regarding data from our deployment study.}

\subsection{Places and people entered in the System}\label{appendix:places}

\begin{table*}[h]
\sffamily\small
\def\arraystretch{1.25}\setlength{\tabcolsep}{0.3em}
\centering

\caption{Regularly visited places and frequently interacted-with people entered in the system. Note that we report places and peoples in simplified categories by anonymizing them.}
\Description{A table listing regularly visited places and frequently interacted-with people for 10 dyads (D1--D10). The first column shows dyad identifiers; the second column lists places such as schools, welfare centers, after-school centers, therapy clinics, sports clubs, and various classes (e.g., swimming, badminton, art, cooking); the third column lists people categories with counts in parentheses, including family members (mother, father, grandmother), teachers (ranging 3--14), and friends (ranging 2--5). Places and people are anonymized into simplified categories. D4 and D5 show the most diverse locations and highest friend counts (25 and 21), while D9 and D10 have fewer entries overall.}
\label{tab:places}
\begin{tabular}{%
|c!{\color{lightgray}\vrule}
m{0.55\textwidth}!{\color{tablegrayline}\vrule}
m{0.37\textwidth}|}
\hline

\rowcolor{tableheader}
\textbf{Dyad} & \textbf{Regular visited places} & \textbf{Frequently interacted-with people} \\

\arrayrulecolor{black}\hline

\textbf{D1} & School, Welfare center, Swimming lesson, Community, Taekwondo class & Teachers (5), friends (3)
\\ \arrayrulecolor{tablegrayline}\hline 
\textbf{D2}    & School, center, home & Mother, father, grandmother, teachers (5)
\\ \hline
\textbf{D3} & School, welfare center, autism clinic, group therapy, church, park & Mother, father, evangelist, teachers (6), friends (9) 
\\ \hline
\textbf{D4}    & School, language center, after-school center, welfare center, cooking class, art class, clarinet class, drone class, badminton class, soccer club,\newline{}bowling club & Mother, bowling club staff (3), teachers (12),\newline{}friends (25)
\\ \hline
\textbf{D5}    & School, developmental PE center, language center, cognitive center, \newline{}after-school center, welfare center, badminton class, social skills class,\newline{}cooking class, drone class, piano academy, art academy, sports club
    & Mother, father, teachers (14), friends (21) 
\\ \hline
\textbf{D6}   & School, after-school center, PE class, badminton class, swimming class, piano lesson, art class, workbook study, SNPE exercise & Mother, teachers (14), friends (13)
\\ \hline
\textbf{D7}    & School, welfare center, English academy, badminton class, swimming class, social skills PE class, art class, cooking class 
& Teachers (10), friends (14)
 \\ \hline
\textbf{D8}    & School, game chatroom, hospital, church, apartment garden, \newline{}home visit class & Mother, aunt, cousin, apartment resident (2),\newline{}teachers (7), friends (5)
\\ \hline
\textbf{D9}    & School, after-school class, group exercise class, development center &
    Mother, teachers (4), friends (2)
\\ \hline
\textbf{D10}   & School, counseling center, study room, art class, badminton class & Teachers (4), friends (2)
 \\ \arrayrulecolor{black}\hline                      
\end{tabular}%
\end{table*}
\newpage
\subsection{Categorization of Journal Topics}\label{appendix:categories}
\begin{table*}[h]
\sffamily\small
\centering
\def\arraystretch{1.2}\setlength{\tabcolsep}{0.3em}
\caption{Three topic categories and 13 activity types, number of journal entries and adolescent participants, and example titles from journal entries.}
\Description{The table organizes adolescents' journal topics into three categories—Leisure (56; 46\%), Daily Life (42; 34\%), and Special Activities (24; 20\%)—and lists 13 activity types nested under these categories. Columns report the activity type, the number of journal entries with its percentage of all entries (e.g., Sports \& physical activities 29 (23.77\%)), the number of adolescent participants who wrote at least one entry in that type (range 2--8), and example journal titles with the writer's alias (C1--C10). Within Leisure, the most frequent types are Sports \& physical activities (29 entries, 8 participants), followed by Eating out (10; 5 participants), Artistic pursuits (9; 5), and Playing video games (8; 5). Within Daily Life, School activities leads (12; 6), then Domestic cooking/eating (8; 6), Daily activities (7; 6), Health and well-being (5; 3), Issues and conflicts (4; 3), Learning (4; 3), and Autism therapy (2; 2).
Within Special Activities, Outings accounts for most entries (22; 8), while Competition appears less often (2; 2). Example titles illustrate typical content (e.g., "Joy of a Successful Bowling Turkey – C4," "Buffet with Mom and Dad – C2," "Excited on the First Day of School – C3," "Holiday Trip with [Friend] – C6"). No personal names are revealed; adolescent IDs are anonymized as C1--C10.}
\label{tab:journal_activity}
\begin{tabular}{|m{0.12\textwidth}m{0.23\textwidth}!{\color{lightgray}\vrule}cc!{\color{lightgray}\vrule}m{0.35\textwidth}|}
\hline
\rowcolor{tableheader}

\textbf{Categories} & \textbf{Activity types} & \textbf{Journal count} & \textbf{Participants} & \textbf{Journals} \\
\hline

\multirow{4}{*}{\parbox{0.2\textwidth}{\textbf{Leisure}\newline{}56 (46\%)}} 
  & Sports \& physical activities         & 29 (23.77\%) & 8 & \textit{Joy of a Successful Bowling Turkey} - \child{4}\newline{}\textit{Fun with Badminton} - \child{5} \\\arrayrulecolor{tablegrayline}\cline{2-5}
  & Eating out      & 10 (8.20\%) & 5 & \textit{Buffet with Mom and Dad} - \child{2}\newline{}\textit{Ate Pasta with Mom} - \child{7} \\\cline{2-5}
  & Artistic pursuits            & 9 (7.38\%)  & 5 & \textit{Making a Cloud Slime} - \child{2}\newline{}\textit{[Adolescent]'s Clarinet Practice} - \child{4} \\\cline{2-5}
  & Playing video games           & 8 (6.56\%)  & 5 & \textit{Playing Brawl Stars with [Friend]} - \child{1}\newline{}\textit{Shooting Game with [Friend]} - \child{8} \\
\arrayrulecolor{darkgray}\hline

\multirow{6}{*}{\parbox{0.2\textwidth}{\textbf{Daily Life}\newline{}42 (34\%)}}
  & School activities & 12 (9.84\%) & 6 & \textit{Excited on the First Day of School} - \child{3}\newline{}\textit{Story from [Teacher]'s Class} - \child{10} \\
  \arrayrulecolor{tablegrayline}\cline{2-5}
  & Domestic cooking/eating & 8 (6.56\%)  & 6 & \textit{Happy Family Dinner Time} - \child{7}\newline{}\textit{Making Shaved Ice with [Friend]} - \child{4} \\\cline{2-5}
  & Daily activities & 7 (5.74\%)  & 6 & \textit{Walk with Mom} - \child{10}\newline{}\textit{Happy Movie Time with Dad (At Home)} - \child{7} \\\cline{2-5}
  & Health and well-being          & 5 (4.10\%) & 3 & \textit{Went to Hospital for Stomachache} - \child{8}\newline{}\textit{[Adolescent]'s Sick Day with a Cold} - \child{6} \\\cline{2-5}
  & Issues and conflicts          & 4 (3.28\%)  & 3 & \textit{[Friend] Felt Upset} - \child{3}\newline{}\textit{Counseling with [Teacher]} - \child{10} \\\cline{2-5}
  & Learning        & 4 (3.28\%) & 3 & \textit{Korean Class with [Teacher]} - \child{9}\newline{}\textit{[Adolescent] Studying English} - \child{7} \\\cline{2-5}
  & Autism therapy & 2 (1.64\%) & 2 & \textit{[Adolescent]'s Class at the Welfare Center} - \child{1} \\
\arrayrulecolor{darkgray}\hline
\multirow{2}{*}{\parbox{0.1\textwidth}{\textbf{Special\newline{}Activities}\newline{}24 (20\%)}} 
  & Outings         & 22 (18.03\%) & 8 & \textit{Holiday Trip with [Friend]} - \child{6}\newline{}\textit{Fun Day at Amusement Park} - \child{5} \\\arrayrulecolor{tablegrayline}\cline{2-5}
  & Competition     & 2 (1.64\%) & 2 & \textit{[Adolescent] Won a Prize} - \child{5} \\\arrayrulecolor{black}\hline
\end{tabular}
\end{table*}


\end{document}